\documentclass[aps,prb,twocolumn,superscriptaddress]{revtex4} 

\usepackage{graphics} 
\usepackage{graphicx}
\usepackage{amsmath}
\usepackage{amsfonts}
\usepackage{dcolumn}
\newcommand{\Sh}{Schr\"odinger{ }}
\newcommand{\etal}{\emph{et al.}}
\newcommand{\mca}[1]{\mathcal{#1}}
\newcommand{\bscal}[1]{\bs{\mathcal{#1}}}
\newcommand{\bs}[1]{\boldsymbol{#1}}
\newcommand{\mb}[1]{\mathbb{#1}}
\newcommand{\Fig}[1]{Fig. \ref{#1}}
\newcommand{\braket}[3]{\left\langle #1 \left\vert #2
            \right\vert #3 \right\rangle}
\newcommand{\brakete}[3]{\left\langle #1 \right\vert #2
            \left\vert #3 \right\rangle}
\newcommand{\brak}[2]{\left\langle #1 \left\vert
             #2 \right. \right\rangle}
\newcommand{\ket}[1]{\left\vert #1 \right\rangle}
\newcommand{\fuve}[1]{\left(\bs{#1}\right)}
\newcommand{\fues}[1]{\left(#1\right)}
\newcommand{\fuel}[2]{\fues{#1,\bs{#2}}}
\newcommand{\fuen}[3]{\fues{\bs{#1};#2,\bs{#3}}}
\newcommand{\fusa}[3]{_{\alpha,\bs{#1}}\fues{#2,\bs{#3}}}
\newcommand{\fuap}[3]{_{\alpha\pr,\bs{#1}}\fues{#2,\bs{#3}}}
\newcommand{\yav}[1]{\left[#1\right]}
\newcommand{\llav}[1]{\left\{#1\right\}}
\newcommand{\llal}[1]{\left\{#1\right.}
\newcommand{\abs}[1]{\left\vert#1\right\vert}
\newcommand{\num}{\nonumber}
\newcommand{\non}{\nonumber \\}
\newcommand{\delt}[1]{\delta\left(#1\right)}
\newcommand{\deltus}{\delta\left(\bs{k}\pr-\bs{k}\right)}
\newcommand{\Eq}[1]{Eq. (\ref{#1})}
\newcommand{\Eqs}[1]{Eqs. (\ref{#1})}
\newcommand{\rep}[1]{(\ref{#1})}
\newcommand{\pr}{^{\prime}}
\newcommand{\nas}[1]{\nabla_{\bs{#1}}}
\newcommand{\sa}{_{\alpha}}
\newcommand{\sap}{_{\alpha\pr}}
\newcommand{\sak}{_{\alpha,\bs{k}}}

\begin{document}  

\title[]{ Quantum Hall effect beyond the linear response approximation }   
\author{Alejandro Kunold } 
\email[ Email:]{akb@correo.azc.uam.mx} 
\affiliation{Departamento de Ciencias B\'asicas, Universidad Aut\'onoma
Metropolitana-Azcapotzalco, Av. San Pablo 180,  M\'exico D. F. 02200, M\'exico}  
\author{Manuel Torres}
\email[Email:]{torres@fisica.unam.mx} 
\affiliation{Instituto de F\'{\i}sica,
Universidad Nacional Aut\'onoma de M\'exico, Apartado Postal
20-364,  M\'exico Distrito Federal 01000,  M\'exico}  

\date{\today}  

\begin{abstract}
The  problem  of  Bloch electrons  in two dimensions subject to  magnetic and
intense electric fields is investigated, the
quantum Hall conductance is calculated beyond the linear response approximation. 
Magnetic  translations, electric evolution  and energy translation operators are
used to specify the solutions of the Schr\"odinger
equation. For rational values of  the  magnetic flux quanta per unit cell and
commensurate  orientations of the electric field relative to the original lattice,
an extended superlattice is defined and a  complete set  of mutually commuting
space-time symmetry operators are obtained.  Dynamics of the system is  governed
by   a finite difference  equation that exactly includes the  effects of:
an arbitrary periodic potential,  an electric field orientated in a commensurable
direction of the lattice,  and  coupling between Landau levels.
A  weak  periodic potential broadens  each Landau level in a series of minibands,
separated by the corresponding minigaps; additionally the effect of the electric
field in the energy spectrum is to superimpose  equally spaced discrete levels,
in this ``magnetic Stark ladder"  the energy separation is an integer multiple
of  $ h E / a B $, with $a$ the lattice parameter.
A closed expression for the Hall conductance, valid to all
orders in $\bs E$ is obtained, the leading order term reduces
to the result of Thouless \etal , in which $\sigma_H^{(0)}$
is quantized in units of $e^2/h$. The first order corrections
exactly cancels  for any miniband that is completely filled.
Second order corrections for the miniband conductance are explicitly
calculated as $\sigma_H^{(2)} \propto e^3/ U_0^2 B$, with $U_0$ the
strength of the periodic potential. However the use of a sum rule shows that 
$\sigma_H^{(2)}$ cancels when a Landau band is fully occupied.  
\end{abstract}
\maketitle 

\section{Introduction}\label{intro}

The  quantization of the Hall conductance ($\sigma_H$) in integer \cite{Klit1}
and fractional values \cite{Tsui1} is an astonishing and unexpected result,
because in a real sample  there are many effects that would be expected
to flaw this possibility. In  the  integer quantum Hall effect (IQHE),
the conductance is an integer multiple   of $e^2/h$; simultaneously the
longitudinal magnetoresistance vanishes; both results are observed
within an experimental uncertainty of less than $3 \times 10^{-8}$. 
For the IQHE  the exact quantization of the  Hall conductivity was
explained based on a topological reasoning  by  Thouless
\etal \cite{Thou1,Thou2,Khomo1}; $\sigma_H$ in units of $e^2/h$ is a
topological invariant, the so-called first Chern number  \cite{Avron1,Simon1},
which can only take integer values. $\sigma_h$ is an integer multiple
of $e^2/h$ if the Fermi level lies in a gap between Landau levels. 
The addition of a  periodic potential broadens  each Landau level in
a series of minibands, separated by the corresponding minigaps.
Contrary  to the obvious suggestion that each  subband carries a
fraction of $e^2/h$,  Thouless \etal demonstrated that each 
filled miniband  also contributes with an integer multiple of $e^2/h$
to $\sigma_H$. The Hall conductance varies in a non-monotonous sequence
as the Fermi level sweeps through contiguous minigaps.
Clear experimental evidence for the internal structure of the
Hofstadter butterfly  spectrum has only been found very recently
in the measurement of  $\sigma_H$ for lateral superlattices \cite{Klit2}. 
The work of Thouless \etal  makes use of the Kubo linear response theory,
it is of our  interest to consider the structure of the Hall current
when a finite electric field ($\bs E$) is applied, in particular we
shall be able to compute higher order corrections on $\bs E$ to the
Hall conductivity. 

The problem of electrons moving under the  simultaneous influence of a periodic
potential and  a magnetic field has been discussed by many
authors \cite{Peierls1,Harper1,Zak1,Azbel1,Rauh1,Dana1,Harper3,Zak2}; the 
spectrum displays an amazing complexity including various kinds of
scaling and a Cantor set structure \cite{Hofta1}. Some of these results
had been used in order   
to disclose the topological structure of  the Hall conductance within the
linear response theory.
The addition of the electric field leads to the possibility of testing the
exactness of the
quantization  beyond the linear approximation. Furthermore the problem
by itself possesses a rich 
physical structure that makes  its analysis worthwhile. 

We consider the problem of an electron moving in a two-dimensional
lattice in the presence of applied  magnetic
and electric  fields. We refer to this as the electric-magnetic
Bloch problem (EMB).
The corresponding magnetic Bloch system  (MB) has a long and rich
history.
An early important contribution was made by Peierls \cite{Peierls1}
who  suggested the substitution of the  Bloch index $\bs k$ by the
operator $\left(\bs p - e \bs A \right)$ in the $B = 0$ dispersion
relation  $\epsilon(\bs k)$, which is then treated as an effective
single-band Hamiltonean.
The symmetries of the MB problem were analyzed by Zak \cite{Zak1,Dana1,Zak2},
who worked out the representation
theory of the  of magnetic-translations group. The renowned Harper equation
was derived assuming a tight-binding approximation in which  the magnetic field
acts as a perturbation that splits the Bloch bands  \cite{Harper1}.
Rauh derived a dual Harper equation in the opposite limit of intense magnetic
field \cite{Rauh1},
here the periodic potential acts as a perturbation that broadens the Landau levels,
in this case the Harper equation takes the form  
\begin{equation}\label{harp0}
c_{m-1}+2\cos\fues{2\pi \sigma m + \kappa} c_{m}
+ c_{m+1}=\varepsilon c_{m}.
\end{equation}
where  $\sigma = 1/\phi$ is the  inverse of magnetic flux $\phi$ in a cell
in units of $h/e$.  The studies
of the butterfly spectrum by  Hofstadter and
others   have since created an  unceasing  interest in the problem
because of the beautiful self-similar structure of the
butterfly spectrum \cite{Hofta1}.  Remarkably, an experimental realization of the
Hofstadter butterfly  was  achieved  considering  the transmission of
microwaves \cite{Kuhl1} and
acoustic waves  \cite{Richoux1} through an array of scatterers.

The symmetries of the EMB problem were analyzed by Ashby and Miller,
who constructed the group of the electric-magnetic translation
operators, and worked out their irreducible representations \cite{Ashby1}.
The properties of the electric-magnetic operators were utilized in order
to derive a finite difference equation that governs the dynamics of the
EMB problem when  the coupling between Landau levels can be
neglected \cite{Kunold1}. In this paper we shall derive the
equation that applies under most general conditions.
Magnetic  translations, electric evolution and energy translation
operators are used to specify the solutions of the Schr\"odinger
equation, commensurability conditions must be implemented in order
to obtain a set of mutually commuting space-time symmetry
operators. In addition to the broadening of the
Landau levels induced by the  periodic potential, the effect of the
electric field in the energy spectrum is to superimpose
equally spaced discrete levels; in this ``magnetic Stark ladder"
the energy separation is an integer multiple of $h E/aB$,
with $a$ the lattice parameter. A closed expression for the Hall
conductance, valid to all order in $\bs E$ is obtained. 
The leading order contribution is quantized  in units of $e^2/h$.
First and second order correction are explicitly computed and shown
to take the form: $\sigma_H^{(1)}=0$ and
$\sigma_H^{(2)}\propto e^3/ U_0^2 B$ respectively, for any miniband that 
is completely filled.

The paper is organized as follows. In the next section we present
the model that describes the EMB problem  and  construct its symmetry
operators. In Section \ref{wave} we describe the commensurability 
conditions required to have simultaneously commuting operators, they
are exploited in order to construct  an appropriated wave function basis.
We derive an effective equation \rep{pro3}, in which the
``evolution" is determined by  a differential equation  with respect to
the longitudinal pseudomomentum, this equation becomes essential in order
to obtain a closed expression for the Hall conductance \rep{xyz1}.
The derivation of the finite difference equation that governs the
dynamics of the system is presented in Section \ref{harper}.
Results for the  energy spectrum are presented, and are also
discussed from the perspective of the adiabatic approximation. 
A closed expression for the Hall conductance, valid to all
order in $\bs E$ is obtained in \ref{hall}, the leading
order term $\sigma_H^{(0)}$ and perturbative corrections
$\sigma_H^{(1)}$ and  $\sigma_H^{(2)}$ are calculated. 
The last section contains a summary of our main results. 

\section{The electric-magnetic Bloch problem}\label{embp}
\subsection{The model}\label{model}
  
Let us consider the  motion of an electron in a two-dimensional
periodic potential,  subject  to a uniform magnetic field $B$
perpendicular  to the plane and to a constant electric field $\bs E$,
lying on the plane according to  $\bs E =  E (\cos \theta, \sin \theta)$
with $\theta$ the angle between  $\bs E$ and the lattice 
$x-$axis. The dynamics of the electron is governed by a time-dependent
Schr\"{o}dinger equation that for convenience is written as 
\begin{equation}\label{shr1}
S\ket{\psi}=
\yav{\frac{1}{2 m^*}\fues{\Pi_x^2+\Pi_y^2}+V-\Pi_0}
\ket{\psi}=0,
\end{equation}
here  $m^*$ is the effective  electron mass,  $ \Pi_{\mu}=p_{\mu}+eA_{\mu}$, 
with  $p_{\mu}= (\imath \hbar \partial/\partial t, - \imath \hbar \nabla )$.
Except for the final results for the conductivity, throughout the paper energy
and  lengths are  measured in units of 
$ \hbar\omega_c=\frac{\hbar e B}{m^*},$ and $ \ell_0=\sqrt{\frac{\hbar}{e B}}$,
respectively, where  $\omega_c$ is the cyclotron frequency and $ \ell_0$
the magnetic length.  So, unless specified, we set
$\hbar=e=m^*=1$; although $B$ can also be omitted from the expressions,
we find convenient to  explicitly display it.   Covariant notation will
be used  to simplify the expressions, $e.g.$
$x_{\mu}=\fues{t,\bs{x}}=\fues{t,x,y}$.

Equation  (\ref{shr1}) can be considered as an eigenvalue equation for the operator 
$S$ with eigenvalue $0$.  The gauge potential is written  in an arbitrary gauge,
it includes two gauge parameters  $\alpha$, and  $\beta$,
the final physical results should be of course, independent of the gauge. 
Hence the components of the gauge potentials are written as 
\begin{align}
A_0&=  \fues{ \beta- \frac{1}{2} } \, \bs{x} \cdot \bs{E}
,\non
A_x&=-\fues{\beta +\frac{1}{2}}E_xt
+\fues{\alpha -\frac{1}{2}} B y,\label{pve0}\\
A_y&=-\fues{\beta +\frac{1}{2}}E_yt
+\fues{\alpha +\frac{1}{2}} B x \, . \num
\end{align}
The symmetrical gauge is recovered for $\alpha = 0$, whereas the Landau
gauge corresponds to the selection
$\alpha = 1/2$. If $ \beta= 1/2$ all the electric field contribution appears
in the vector potential, instead  if 
$ \beta= - 1/2$ it lies in  the scalar potential and the \Sh equation becomes time 
independent, however even in this case a time dependence will slip  into the problem
through the symmetry operators.  A general two-dimensional periodic potential
can be represented in terms of  its Fourier decomposition
\begin{equation}\label{pot0}
V\fues{x,y}=\sum_{r,s}v_{rs}
\exp\fues{\imath \frac{2\pi  r x}{a}+\imath \frac{2\pi s y}{a}} \, . 
\end{equation}
For specific numerical results we shall use  the potential
\begin{equation}\label{pot1}
V\fues{x,y}=U_0\yav{\cos\fues{\frac{2\pi x}{a}}
+\lambda\cos\fues{\frac{2\pi y}{a}}}.
\end{equation}
$\lambda$ is a parameter that can be varied in order to have an anisotropic
lattice; $\lambda=1$ corresponds to the isotropic limit. 

\subsection{Electric evolution and magnetic  translations}\label{magtrans}

Let $(t, \bs x) \to (t + \tau , \bs x + \bs R)$  be
a uniform translation in space and time, where $\tau$ is an
arbitrary time and $\bs R$ is a lattice vector.
The classical equations  of  motion remain  invariant under these
transformations; whereas  the  Schr\"{o}dinger equation does not,
the reason being the space and time dependence of the gauge potentials.
Nevertheless,  quantum dynamics  of the system  remain invariant under
the combined  action of  space-time translations  and  gauge
transformations. The electric and magnetic translation 
operators are defined as
\begin{equation} \label{eqmod3}
T_0(\tau) = \exp{ (- \imath \tau {\cal O}_0)}  \, , \qquad
 T_j (a) =   \exp{(\imath a \,  {\cal O}_j )}  \, , 
\end{equation}
with  $j = x,y$ and the  electric-magnetic symmetry generators  are written as
covariant derivatives $ {\cal O}_\mu = p_\mu +   \Lambda_\mu$,  
with the components of the dual gauge potentials $ \Lambda_\mu $ given by
\begin{align}\label{dualgauge}
 \Lambda_0&= A_0 +  \bs x \cdot \bs E \, ,
\non
 \Lambda_x&= A_x + B y + E_x t    \,,   \qquad 
 \Lambda_y  = A_y  - B x + E_y t  \, .  
\end{align}
It is straightforward to prove that  the operators in (\ref{eqmod3})
  are indeed symmetries of the \Sh equation; 
they commute with the operator $S$  in Eq. (\ref{shr1}).   Similar expressions
for the electric-magnetic operators were given by Ashby and Miller \cite{Ashby1},
however their definition  included  simultaneous  space and time translations;
we deemed it  more convenient  to separate the effect of the time evolution
generated by  the    $T_0 $ to that of the space translations generated by
$ T_j$. The following commutators can be worked out 
\begin{align}\label{conmu1}
 \left[\Pi_0,\Pi_j \right] = & -\imath E_j  \, ,  \qquad \qquad
 \left[\Pi_1,\Pi_2 \right] = -\imath B \, , 
\nonumber \\
\left[ {\cal O}_0,  {\cal O}_j\right] = &  \,\, \imath E_j \, ,
\qquad \qquad \,\,\,\,\,
\left[ {\cal O}_1, {\cal O}_2 \right] =\imath B \, , \nonumber \\
\left[\Pi_\mu, {\cal O}_\nu \right] = & \,\, 0  \, .
\end{align}
Notice that the electric-magnetic generators   ${\cal O}_\mu$ have been defined
in such a form that they commute with all the velocity operators   $\Pi_\nu $.    
The Schr\"{o}dinger equation and the symmetry operators are
expressed in terms of covariant derivatives
$\Pi_\mu$ and ${\cal O}_\mu$, respectively.
A dual situation in which the roles of $\Pi_\mu$ and ${\cal O}_\mu$
are interchanged can  be considered. The dual problem corresponds
to a simultaneous reverse in the directions of $B$ and $\bs E$.
We notice that the commutators in  the second line of Eq. (\ref{conmu1})
are part  of the  Lie  algebra of the EM-Galilean two dimensional
group \cite{Hadji1,Hadji2}. This group is obtained when  the usual rotation
and  boost operators of  the planar-Galilean group  are replaced  by their
electric-magnetic  generalization in which the operators  are enlarged by
the effect of a gauge transformation. 

 \section{Electric-magnetic Bloch Functions}\label{wave}
\subsection{Commensurability conditions}\label{commen}

In order to construct a complete base that  expands the wave function we 
require the symmetry operators to  commute with each other. However we have
\begin{equation}\label{nco1}
T_{\mu}T_{\nu}=
e^{c^{\mu}c^{\nu}\yav{\mca{O}_{\nu},\mca{O}_{\mu}}}
T_{\nu}T_{\mu} \, . 
\end{equation}
A set of simultaneously commuting symmetry operator can be found if appropriated
commensurate conditions  are imposed, we follow a   three   step method to find them:
\begin{enumerate}
\item First  we consider a  frame rotated  at  angle $\theta$,   with axis along
the longitudinal and transverse direction relative to the electric field.
An orthonormal basis for   this frame is  given by
$\bs{e}_L = (cos \theta, sin \theta)$,  $\bs{e}_T = (-  sin \theta, cos \theta)$
and $\bs{e}_3=\bs{e}_L \times\bs{e}_T$. The electric field is parallel to
$\bs{e}_L$ and the magnetic field points along $\bs{e}_3$. 
We assume a   particular orientation of  the electric field,
for which  the following condition holds
\begin{equation}\label{rho1}
\tan \theta=\frac{E_y}{E_x}=\frac{m_2}{m_1}
\end{equation}
where  $m_1$ and $m_2$   are relatively prime integers.
This condition insures that spatial periodicity is also found
both along the transverse and the  longitudinal directions.
Hence, we define a rotated lattice spanned by the longitudinal
$\bs b_L = b \bs{e}_L $ and transverse  $\bs b_T = b \bs{e}_T$
vectors, where $b=a\sqrt{m_1^2 +  m_2^2}$.
The  spatial components of the symmetry generator $\bscal O$ are
projected along the longitudinal and transverse directions:
$\mca{O}_L = \bs{e}_L \cdot \bscal{O}$ and
$\mca{O}_T  = \bs{e}_T \cdot \bscal{O}$.
It is readily  verified that $ \yav{\mca{O}_0,\mca{O}_T}=0$.

\item  For the rotated lattice,  we regard  the  number of flux quanta per
unit cell  to be a rational  number  $p/q$, that is 
\begin{equation}\label{sig1}
\phi \equiv {1 \over \sigma}={B \, b^2 \over 2\pi }={p \over q} \, . 
\end{equation}
We can then define a extended superlattice. A rectangle made of $q$
adjacent lattice cells of side $b$ contains an integer number of  flux quanta.
The basis vectors of the superlattice are chosen as $q \bs b_L$ and $\bs b_T$.
Under these conditions the longitudinal and transverse magnetic translations
$T_L \fues{qb}=\exp{(\imath q b {\cal O}_L)}$ and
$T_T\fues{b}= \exp{( \imath b {\cal O}_T)}$
define commuting symmetries under displacements $q\bs b_L$ and $\bs b_T$.  
Henceforth we shall either use the subindex $(L,T)$ or $(1,2)$
to label the longitudinal and transverse directions. 

\item We observe that  $T_0$ and  $T_L(q b)$ commute with 
$T_T$.  Yet they fail to  commute with each other: 
\begin{equation}\label{tau11}
T_0\fues{\tau }T_L \fues{qb}=
e^{-\imath qb\tau E}
T_L\fues{qb}T_0\fues{\tau}.
\end{equation}
However  the  operators $T_0$ and  $T_L(q b)$  will  commute with one another
by  restricting   time,   in the   evolution operator,  to  discrete values
with period 
\begin{align}
\tau&=n\tau_0,& n&\in \mathbb{Z},&
\tau_0&=\frac{2\pi}{qbE}=
\frac{1}{p}\fues{\frac{b}{v_D}}\label{tau1} \, , 
\end{align}
where the drift velocity is $v_D =  {E / B}$  and  we utilized Eq. (\ref{sig1}) to
write the second equality.
$b/v_D$ is the period of time it takes an electron with  drift velocity $v_D$ to
travel between lattice points. \Eq{tau1} can be interpreted as the condition that
the ratio of  two energy scales is an integer: as we shall see ($\sigma bE$) is
the magnetic-Stark ladder spacing, whereas $(2 \pi v_D /b)$ is the quasienergy 
Brillouin width;  hence Eq. (\ref{tau1}) represents  the ratio of these two quantities.
\end{enumerate}

We henceforth consider that the three conditions (\ref{rho1}),  (\ref{sig1}), and
(\ref{tau1}) hold simultaneously. In this case the three EM operators:
the electric evolution ${\cal T}_0 \equiv T_0(\tau_0)$, and the magnetic translations
${\cal T}_L \equiv T_L(q b)$, and ${\cal T}_T \equiv T_T(b)$ form a  set of mutually
commuting symmetry operators. In addition to the symmetry operators it is convenient
to define the  energy translation operator \cite{Zak3}
\begin{equation}\label{newop}
{\cal  T}_E \,    =  \exp{ \left(-\imath {2 \pi \over \tau_0} t  \right) }   \, , 
\end{equation}
that produces a finite translation in energy by ${2 \pi /  \tau_0} \equiv q b E$.
${\cal  T}_E $ commutes with the three symmetry  operators but not with $S$.
Its eigenfunctions
\begin{equation}\label{quasit}
\mca{T}_E   \psi = e^{\imath qbE\vartheta} \psi,
\end{equation}     
define a quasitime  $\vartheta$  modulo  $\tau_0$. 

\subsection{Wave function and generalized Bloch conditions}\label{blochcond}

Having defined  ${\cal  T}_0$, ${\cal T}_L$ and
${\cal T}_T$ that commute with each other and  also
with $S$,  it is possible to seek for solutions of the
Schr\"{o}dinger  equation labeled by the quasienergy
(${\cal E}$)  and the longitudinal  ($k_1$) and transverse
($k_2$) quasimomentum according to
\begin{align}
\mca{T}_0 \ket{\mca{E},k_1,k_2} &=e^{-\imath \tau_0\mca{E}}
\ket{\mca{E},k_1,k_2}, \non
\mca{T}_L\ket{\mca{E},k_1,k_2}  &=e^{\imath k_1qb}
\ket{\mca{E},k_1,k_2}, \label{trf1} \\
\mca{T}_T \ket{\mca{E},k_1,k_2} &=e^{ \imath k_2b}
\ket{\mca{E},k_1,k_2}.\num
\end{align}
The magnetic Brillouin zone (MBZ) is defined by 
$k_1\in\yav{0,2\pi/qb}$  and $k_2\in\yav{0,2\pi/b}$.
Similarly the quasienergy  $\mca{E}$
is defined modulo $2\pi/\tau_0=qbE$. If a restricted energy scheme is selected 
for the energy,  the first energy Brillouin region is defined by the condition
$\mca{E}\in\yav{0,2\pi/\tau_0}$.
We shall find convenient to perform a  canonical transformation to new variables
according to
\begin{align}
Q_0&=t,&
P_0&=\mca{O}_0 -\frac{1}{2}E^2,\non
Q_1&=\Pi_2+ E,&
P_1&=\Pi_1 ,\label{ops1}\\
Q_2&=\mca{O}_1 - E t,&
P_2&=\mca{O}_2+E ,\num
\end{align}
that satisfy the commutation rules
$\yav{Q_{\mu},P_{\nu}}=-ig_{\mu\nu}, \quad  g_{\mu\nu}= Diag(-1,1,1)$.
Applied to Eq. (\ref{shr1}) the transformation yields for the \Sh equation 
\begin{align}\label{shr3}
P_0  \ket{ \psi }&=  H  \ket{ \psi}   \, ,  \non
  H &=  \yav{\frac{1}{2} \fues{P_1^2+Q_1^2} +
V\fues{x,y} -E  P_2 } \, .
\end{align}
In this equation  $(x, y)$ must be expressed in terms of the new variables:
\begin{align}\label{relxQ}
{x \over a} \,=\,  & {m_1 \over b} \, \left( Q_1 - P_2 \right) - {m_2 \over b}
\, \left( Q_2 - P_1 \right)  \, , \num \\
{y \over a} \,=\, & {m_2 \over b} \, \left( Q_1 - P_2 \right)  + {m_1 \over b}
\, \left( Q_2 - P_1 \right)  \, .
\end{align}
On the other hand,  the EM-symmetry operators and the 
energy translation operator take the form:
\begin{align}
\mca{T}_0 &=e^{-\imath \tau_0P_0}, \qquad \qquad 
\mca{T}_L =e^{\imath qb\fues{Q_2+EQ_0}},\label{cco1}\\
\mca{T}_T &=e^{\imath  bP_2},  \qquad \qquad 
{\cal  T}_E \,   =  e^{ -\imath  {2 \pi \over \tau_0} Q_0   }   \, .  \num
\end{align}
Notice that these operators do not depend on the variables $P_1, Q_1$. Thus, 
it is  natural to  split the  phase space $(Q_\mu,P_\mu)$ in the  $(Q_1,P_1)$
and the $(Q_0,P_0;Q_2,P_2)$ variables. For the first set   a harmonic oscillator
base is used. Whereas for the subspace generated by the variables
$(Q_0,P_0;Q_2,P_2)$, we observe that the  operators
(${\cal  T}_0^i$, ${\cal T}_L^j$, ${\cal T}_T^k$,  ${\cal T}_E^l$)
with all possible integer values of $(i,j,k,l)$ form a complete
set of operators.
The demonstration follows similar steps as those presented by
Zak in reference \cite{Zak4}.  Hence a complete set of functions,
for the subspace $(Q_0,P_0;Q_2,P_2)$, is provided by the eigenfunctions
of the operators (${\cal  T}_0^i$, ${\cal T}_L^j$, ${\cal T}_T^k$,
${\cal T}_E^l$). As starting point, we select  a base of eigenvalues
of the following operators
\begin{align}\label{bas5}
A^{\dag}A\ket{\mu,\mca{E},k_2}&=\mu\ket{\mu,\mca{E},k_2},\non
P_0\ket{\mu,\mca{E},k_2}&=\mca{E}\ket{\mu,\mca{E},k_2} ,\\
P_2\ket{\mu,\mca{E},k_2}&=k_2\ket{\mu,\mca{E},k_2},\num
\end{align}
where $A $  and $ A^{\dag}$ are the 
lowering and raising operators of  the $\mu$-Landau level:
\begin{equation}\label{raising}
A= {1 \over \sqrt{2}} (P_1 - \imath Q_1) \, , \qquad
A^{\dag} = {1 \over \sqrt{2}} (P_1 + \imath  Q_1) \, . 
\end{equation}
 It is straightforward to verify that these states 
fulfill the required eigenfunction condition (\ref{trf1}) for  the  $\mca{T}_0$
and $\mca{T}_T$ operators. On the other hand $\mca{T}_L$ induces a shift in 
the $\mca{E}$ and $k_2$ labels, while $\mca{T}_E$ induces a shift
$qbE$ in  the $\mca{E}$ label. The previous considerations suggest that a state 
$\ket{\vartheta,\mca{E},k_1,k_2}$ that is a simultaneous  eigenfunction of the
four operators $\mca{T}_E$, $\mca{T}_0$, $\mca{T}_L$ and
$\mca{T}_T$ with the required eigenvalues   (Eqs. \ref{quasit} and \ref{trf1}),
can be constructed as a linear superposition of states of the form
$ \left[ \mca{T}_L \right]^l  \, \left[   \mca{T}_L  \mca{T}_E^{-1} \right]^m %
\ket{\mu,\mca{E},k_2}$.
We write down the state, and verify their correctness: 
\begin{align}\label{mor3}
& \ket{\vartheta,\mca{E},k_1,k_2}= \non
& \sum_l \yav{\mca{T}_Le^{-\imath qbk_1}}^l
\sum_{\mu,m} c^{\mu}_m\yav{e^{\imath \sigma bQ_2}
e^{\imath \sigma b\fues{E\vartheta-k_1}}}^m
\ket{\mu,\mca{E},k_2} \, . 
\end{align}
It is easy to check that this  function  satisfies the  three  eigenvalue equations
for the symmetry operators  in (\ref{trf1}). 
Additionally it can be verified that  the eigenvalue condition (\ref{quasit}) for
the energy translation operator is also fulfilled 
by  imposing   the periodicity condition $c_{m +p}^\mu =\,  c_m^\mu$.
Every  state in (\ref{mor3}) yields a set of different eigenvalues, a condition that
follows from the fact that 
the selected  set of operators is complete, hence  the base of eigenvalues in
(\ref{mor3})  is complete and orthonormal.   

Up to this point we have constructed a base for the set of four EM translation 
operators, however the operator ${\cal T}_E$ is not a symmetry of the
problem, consequently the  solution of the \Sh
is constructed as a superposition of all the states characterized 
by  $\vartheta$;  the  state becomes 
\begin{align}\label{wav3}
&\ket{\mca{E},k_1,k_2}=\int d\vartheta\;
C\fues{\vartheta}
\ket{\vartheta,\mca{E},k_1,k_2}
= \non
& \sum_{l} \yav{\mca{T}_Le^{-\imath qbk_1}}^l
\sum_{\mu,m}b^{\mu}_m
e^{\imath \sigma b\fues{Q_2-k_1}m}
\ket{\mu,\mca{E},k_2} \, , 
\end{align}
where
\begin{equation}\label{per4}
b^{\mu}_{m}=\int d\vartheta\;
C\fues{\vartheta}c^{\mu}_m
e^{\imath qbE\vartheta m} \, . 
\end{equation}
Equation  \rep{wav3}  represents the correct state that satisfies the eigenvalue
equations (\ref{trf1}). It is convenient to  recast it  in a compact form as 
\begin{equation}\label{wav22}
\ket{\mca{E},\bs{k}}=
\mca{W}\fues{k_1} \ket{\mca{E},k_2} ,
\end{equation}
where  $\ket{\mca{E},\bs{k}}=\ket{\mca{E},k_1,k_2}$,
and the operator  $\mca{W}\fues{k_1}$ is defined as 
\begin{equation}\label{wav21}
\mca{W}\fues{k_1}=\sum_{l}\yav{\mca{T}_Le^{-\imath  qbk_1}}^l
=\sum_{l}e^{\imath qbl\fues{EQ_0+Q_2-k_1}}.
\end{equation}
whereas the ket $\ket{\mca{E},k_2}$ is given by 
\begin{equation}\label{wav25}
\ket{\mca{E},k_2}=\sum_{\mu,m}
e^{\imath \sigma bm\fues{Q_2-k_1}}b_m^{\mu}
\ket{\mu,\mca{E},k_2}.
\end{equation}
The previous expression stresses the fact that  the quantity
$e^{-i\sigma b m k_1} \, \, b_m^{\mu}$ does not depends on $k_1$,
this will be demonstrated below  Eq. (\ref{har3}). These results will
prove to be very useful to analyze the EM problem, in particular they allow
us to obtain an effective \Sh equation where the ``dynamics" is
governed by the derivative with respect to the longitudinal pseudomomentum.
Utilizing \Eqs{wav22} y \rep{wav21} it is straightforward to prove
the following relation 
\begin{align}
 P_0 \ket{\mca{E},k_1,k_2} = &\left[ P_0, \mca{W} \right]  \ket{\mca{E},k_2} +
\mca{W}  P_0  \ket{\mca{E},k_2}
= \non
& \fues{\mca{E}- \imath \bs{E}\cdot\nas{k}}
\ket{\mca{E},k_1,k_2}.
\label{pro2}
\end{align}
Thus the 
\Sh  (\ref{shr3}) can be recast as 
\begin{equation}\label{pro3}
\fues{\mca{E}-\imath \bs{E}\cdot\nas{k}}\ket{\mca{E},k_1,k_2}
=H\ket{\mca{E},k_1,k_2} \, . 
\end{equation}
As it will be lately  discussed,  this form of the \Sh equation becomes essential 
in the discussion of the quantized Hall current. 

The wave function takes a simply form if we adopt the $(P_0,P_1,P_2)$
representation. In this case
$\Psi_{\bs k }\fues{p}=\left\langle P_0,P_1,P_2\vert {\cal E},%
k_1,k_2  \right\rangle$  yields
\begin{align}\label{wf2}
\Psi_{\bs{k}} \fues{P} & =
\sum_{\mu,l,m} \, b_m^\mu \,  \phi_\mu(P_1) 
e^{\imath \left(2 \pi /b \right)  ( m + pl) k_1  } \non
& \delta \left(P_0 - {\cal E} - l q b E \right)
\,\delta\left(P_2- k_{2} + \frac{2\pi}{b}(m + p l ) \right),
\end{align}
where  $\phi_{\mu}\fues{P_1}$ is the harmonic oscillator function  in the 
$P_1$ representation
\begin{equation}\label{hofun}
\phi_{\mu}\fues{P_1}=\brak{P_1}{\mu}=
\frac{1}{\sqrt{\pi^{1/2}2^{\mu}\mu !}}
e^{-P_1^2/2}H_{\mu}\fues{P_1} \, , 
\end{equation}
and  $H_{\mu}\fues{P_1}$  is the Hermite polynomial.

On the other hand,  it is sometimes useful to restore the space-time  representation 
of the wave function.  This problem provides a good example of the use of canonical
transformations  in quantum mechanics. Identifying the matrices that relate the
$(x_\mu, p_\mu)$ variables to $(Q_\mu, P_\mu)$, it is possible to apply the method of
reference \cite{Moch1} in order to obtain the desired transformation. The final
result for the wave function in the space-time  representation  yields
\begin{equation}\label{flo4}
 \Psi_{ \bs{k}}  \left( t, \bs{x} \right)=
e^{\imath \bs{k}\cdot \bs{x}-\imath \mca{E}t} \, \, 
u_{\bs{k}}  \left( t, \bs{x} \right) \, ,
\end{equation}
where the modulation function $u(t,\bs{x}) $ is given by 
\begin{align}\label{wav5}
& u_{\bs{k}} \left( t, \bs{x} \right)  =
\frac{1}{\sqrt{2\pi}}  e^{-\imath x_1 \left[ (\alpha - 1/2) x_2 + k_1 \right]  }
e^{\imath (\beta +  1/2) E t x_1   } \times \non
& \sum_{\mu l m} i^{\mu}  b_{m}^{\mu}
e^{- \imath qbElt }
 e^{\imath \sigma b\fues{k_1 - x_2}\fues{m+pl}} 
 \phi_{\mu}\fues{x_1- k_2  + 2\pi\frac{m+pl}{b}} , 
\end{align}
here  $\phi_{\mu}$ is the same  harmonic oscillator function  \Eq{hofun}, but
now evaluated in the $space$ representation. 

Solution (\ref{flo4}) includes a superposition of Landau-type solutions
originated beneath the spatial and time periodicity.
The spatial periodicity is simply related to the external potential $V$.
Whereas time periodicity arises from the conditions imposed to the symmetry
operators in order to produce commuting symmetries; as discussed in
\Eq{tau1}, the period is given by the time  that takes an  electron
to drift between contiguous lattice points. Notice that \Eq{flo4}
follows from  the  Bloch and  Floquet  theorem. However in the
electric-magnetic case the modulation functions $u\fues{t, \bs{x}}$
are not strictly periodic, instead they satisfy the 
generalized  Bloch conditions 
\begin{align}\label{flo5}
u(t+\tau_0,x_1,x_2)&=
e^{\imath \tau_0 \Lambda_0} \, 
u(t,x_1,x_2),\non
u(t,x_1 +qb,x_2)&=
e^{-\imath qb \Lambda_1} \,
u(t,x_1,x_2),\\
u(t,x_1,x_2+b)&=
e^{-\imath b \Lambda_2} \, 
u(t,x_1,x_2),\num
\end{align}
the phases are determined by the dual-gauge potential that appears
in the symmetry operators \Eq{dualgauge}. It is straightforward 
to verify that $u_{\bs{k}}$ in \Eq{wav5} satisfies these conditions. 
The correct normalization conditions for  the modulation function
are obtained as follows
\begin{equation}
\frac{\fues{2\pi}^2}{qb^2}
\int_{\rm MUC}d^2x\;
u^*\fues{t,x} \, u\fues{t,x}=1 \, , 
\end{equation}
where MUC represents the magnetic cell defined by: 
$x_1\in\yav{0,qb}$ and  $x_2\in\yav{0,b}$.

The function $u$ satisfies conditions similar to those in
\Eq{flo5} but in the MBZ. The MBZ is actually a torus  $T^2$, 
so the edges $(k_1=0,k_2)$ and  $(k_1=2 \pi/ qb,k_2)$ must
be identified as the same set of points, the wave function can differ
at most by a total phase factor (similarly for the edges $(k_1, k_2=0)$
and  $(k_1  , k_2=2 \pi/ b)$):
\begin{align}\label{flo6}
u(k_1 + 2\pi/ qb, k_2) &= e^{\imath f_1} u(k_1,k_2) , \non
u(k_1, k_2+2\pi/ b ) &= e^{\imath f_2} u(k_1,k_2),
\end{align}
the functions $f_1\fuve{k}$ and $f_2\fuve{k}$ are related with
the Hall conductance, however instead of \Eq{flo5}, these functions
are not  analytically known and must be computed numerically. 

\section{Harper generalized equation }\label{harper}
\subsection{Finite difference  equation   }\label{finitedif}

The coefficient  $b^{\mu}_{m}$  in Eq.  \rep{wav3} satisfies a recurrence
relation that is obtained when this base is used to calculate the matrix elements
of the  \Sh equation  \rep{shr3}.  First, let us consider the contribution arising
from the  periodic potential  in \Eq{pot0}. 
The $x$ and $y$ coordinates are written 
in terms of the new variables $(Q_1,P_1,Q_2,P_2)$ by means of Eq.  \rep{relxQ},
producing a term of the form   $\exp \left[ (\imath 2\pi r m_1 (Q_1 - P_2)/b \right]$,
and  another contribution  in which $Q_1$ and $P_2$ are replaced by $Q_2$ and $P_1$.
Once that   $Q_1$ and $P_1$ are replaced by  the raising and lowering operators
in \Eq{raising}, one is lead   to evaluate   the matrix elements of the operator
$D=\exp\fues{z A^{\dag} -z ^* A}$ that  generates  coherent Landau states.
A calculation yields
\begin{align}\label{laguerres}
D^{\nu \mu} & \fues{z}  =\braket{\nu}{\exp\fues{z A^{\dag}-z^*A}}{\mu}
= \non  
&e^{-\frac{1}{2}\abs{ z }^2}
\llal{\begin{array}{ll}\fues{-z^{*}}^{\nu-\mu}
\sqrt{\frac{\nu!}{\mu!}}L^{\mu-\nu}_{\nu}
\fues{\abs{z}^2},& \mu >\nu, \\
z^{\nu-\mu}\sqrt{\frac{\mu!}{\nu!}}
L^{\nu-\mu}_{\mu}\fues{\abs{ z}^2},
& \mu <\nu,\\
\end{array}}
\end{align}
where $L^{\mu}_{\mu}$ are the generalized Laguerre polynomial. 
The  evaluation of the terms that include the $P_2$ operator is direct,
because the base is a  eigenvalue of this operator, on the other hand the
operator $\exp \left[ (\imath 2\pi r m_1 Q_2)/b \right]$ acts as a translation
operator that produces a shift on the index $m$ of the  $b^{\mu}_{m}$ coefficient.
Taking into account these results, it is possible to demonstrate after a lengthly
calculation  that the  \Sh equation  \rep{shr3} becomes 
\begin{equation}\label{idi1}
\sum_{t=-N}^{t=N}  \mb{A}_m^t \fues{k_1,k_2}
{\tilde b} _{m+t}=\fues{\mca{E}+Ek_2+\sigma bEm} {\tilde b}_m \, . 
\end{equation}
Besides its dependence on the index $m$,    ${\tilde b} _m$   is a vector with $L$
components  and  $ \mb{A}_m^t \fues{k_1,k_2}$ is a $L \times L$ matrix   in the
Landau space,    according to 
\begin{align}
{\tilde b_m} &=  \{ b_m^\mu \} =  \fues{b^0_{m},b^1_{m},...,b^L_{m}} \, , \non
\left(  \mb{A}_m^t  \right)^ {\mu \nu} &=
e^{-\imath \sigma bk_1t}\sum_{r,s}B^ {\mu \nu}_{m}\fues{r,s}
\delta_{t,rm_2-sm_1} \, . 
\label{har2}
\end{align}
In the previous expressions    $L$ is the highest  Landau level   included
in the calculations, $N=max \{ (r,s)\fues{m_1+m_2} \}$ is related to the
largest harmonic  in the Fourier decomposition of the periodic potential
\rep{pot0}, and $B^{\nu \mu}_{m}\fues{r,s}$ is given by 
\begin{align}\label{deb1}
& B_{m}^{\nu \mu}  \fues{r,s}= \non
& \llal{\begin{array}{ll}
\fues{v_{00}+\mu+\frac{1}{2}
}\delta^{\mu \nu},& r,s=0,\\
 & \\ v_{rs}D^{\nu \mu}\fues{H_{rs}} e^{\imath K_{rs}}
e^{\imath M_{rs}\yav{\sigma b\fues{rm_2-sm_1+m}+k_2}}&r,s\ne0 , 
\end{array}}
\end{align}
the following definitions where introduced in the previous equation
\begin{align}\label{deb2}
H_{rs}&=\frac{\sqrt{2}\pi}{b}\fues{m_2-\imath m_1}
\fues{\imath r+s},\non
K_{rs}&=\frac{2\pi^2}{b^2}\fues{r m_2-s m_1}
\fues{r m_1+s m_2},\\
M_{rs}&=-\frac{2\pi}{b}\fues{r m_1+s m_2}.\num
\end{align}
We recall that the indexes $(r,l)$ refer to the Fourier expansion of the periodic
potential \Eq{pot0}, whereas $(m_1,m_2)$ correspond  to the integers that
determined the electric field orientation in  \Eq{rho1}. 

Equation  \rep{idi1} describes the system in terms of a recurrence relation on a
two dimensional  $(\mu, m)$ basis.
The coupling in $m$ ranges from $m - N$ to $m + N$.
We follow the method of reference \cite{Risken1}  to recast  the recurrence
relation on $m$ with $N$ nearest-neighbor coupling into a tridiagonal vector
recurrence relation. This is enforced by defining a $N$ component vector 
\begin{equation}\label{relcb}
{\tilde  c}_m=\fues{\begin{array}{ccccc}
{\tilde b }_{Nm-1}, & {\tilde b }_{Nm}, & {\tilde b }_{Nm+1}, & \dots  , 
& {\tilde b }_{N\fues{m+1}-1}
\end{array}} \, , 
\end{equation}
and matrices   $Q^\pm_m$ and $Q_m$  with elements 
\begin{align}
Q^-_m\fues{k_1,k_2}&=\fues{\begin{array}{cccc}
 \mb{A}^{-N}_{Nm} &  \mb{A}^{-N+1}_{Nm} & \dots &   \mb{A}^{-1}_{Nm} \\
0 &  \mb{A}^{-N}_{Nm+1} & \dots &  \mb{A}^{-2}_{Nm+1} \\
\vdots & \vdots & \ddots & \vdots \\
0 & 0 & \dots &  \mb{A}^{-N}_{N\fues{m+1}-1} \non 
\end{array}} \, , \\ \nonumber  \\
Q_m\fues{k_1,k_2}&=\fues{\begin{array}{ccccc}
\mb{A}^{0}_{Nm} &  \mb{A}^{1}_{Nm} & \dots &  \mb{A}^{N-1}_{Nm} \\
\mb{A}^{-1}_{Nm+1} &  \mb{A}^{0}_{Nm+1} & \dots &   \mb{A}^{N-2}_{Nm+1} \\
\vdots & \vdots & \ddots & \vdots  \\
\mb{A}^{1-N}_{N\fues{m+1}-1} &  \mb{A}^{2-N}_{N\fues{m+1}-1} & \dots&
\mb{A}^{0}_{N\fues{m+1}-1}\\
\end{array}}\label{deq1} . 
\end{align}
Using the fact that  $Q^+_m = (Q^-_m)^\dag$,  \Eq{idi1} can be reorganized as a
tridiagonal recurrence relation 
\begin{align}
Q^-_m {\tilde c}_{m-1} & + Q_m {\tilde c}_m+Q^+_m {\tilde c}_{m+1} = \non
& \yav{\fues{\mca{E}+Ek_2+\sigma bEm}I_N+\sigma bED_N} {\tilde c}_m \, . 
\label{har3}
\end{align}
Here, ${\tilde c}_m$ is $N-$dimensional vector according to the definition
\rep{relcb}, additionally each one  of its   components is a $L$
dimensional vector due to its dependence on the Landau index \Eq{har2}.
$I_N$ is the unit matrix and $D_N=Diag \,\, (0,1,2,\dots,N-1)$.
Likewise,  each of the components of the matrices  $Q_m$,  $I_N$,
and  $D_N$ are  $L \times L$ matrices relative to the Landau contributions.
In relation with \Eq{wav25} it was mentioned that the quantity
$ e^{-\imath \sigma b  k_1} \, \,  {\tilde c}_m$ does not explicitly depend
on  the longitudinal  pseudomomentum  component $k_1$, this can be easily
verified if the substitution
${\tilde c}_m^\prime  = e^{-\imath \sigma b  k_1} \, \,  {\tilde c}_m$
is incorporated into \Eq{har3}, then using \Eqs{har2} and \rep{deq1},
it  is straightforward to verify that ${\tilde c}_m^\prime$ is indeed
independent of $k_1$.

\Eq{har3} is one of our main results,  and deserves to be emphasized.
It is a generalization of the Harper equation that exactly  includes
the following  effects:
(1)  an arbitrary periodic potential,
(2) an electric field orientated in a commensurable direction of the
lattice,  and
(3) the coupling between different Landau levels.
So far, no approximations are involved, consequently it holds under
most general conditions. In practice, only a finite number $L$ of Landau
levels can be included on the calculations, but a very good convergence can
be obtained with a reasonable small selection for $L$.
Previous known results are recovered if some approximations are enforced:
($i$) if the electric field is switched off, its orientation is meaningless,
so the integers  $(m_1,m_2)$ can be set to $m_1=1$ and $m_2=0$, in this case
\Eq{har3} reduces to the model previously discussed by
Petschel and Geisel \cite{Geisel1}.
($ii$) If $\bs E = 0$ and additionally the coupling between different
Landau bands  is neglected and the potential is taken as the sum of
cosines given in \Eq{pot1}, then the  system  reduces to a set of Harper
equations, one for each Landau band.
 
Let us return to the case that includes the electric field. Equation \rep{har3}
describes the system dynamics under  most generals conditions; in spite of its
complicated structure the equation can be solved by the matrix continued
fraction methods. However,  a considerable simplification is obtained if  the  electric field is aligned 
along the $x-$lattice axis, i.e. $(m_1=1,m_2=0)$. 
Henceforth, we  consider that the electric field is orientated along the
$x-$axis $(m_1=1,m_2=0)$,  so $b \equiv a$ and additionally  that the 
periodic potential takes the form  given in  \rep{pot1}. The dimension $N$
of the vectors ${\tilde c}_m$  and matrices 
in Eqs. \rep{relcb} and \rep{deq1}  reduces to  $N= max \{ (r,s) (m_1 + m_2\}=1$,
hence the dimensions of  ${\tilde c}_m \equiv c_m$ and   $Q$   reduce  to include
only   the Landau  indexes. Using equations \rep{har2}, \rep{deb1} and \rep{deq1}
the matrices $Q$ take the form 
\begin{align}\label{defQ}
( Q^{+}_m )^{\mu\nu} & \equiv ( Q^{+}  )^{\mu\nu}
=\frac{\lambda \pi K}{2\fues{1+\lambda}a^2} \, e^{\imath\sigma bk_1}
D^{\mu\nu}\fues{\sqrt{\pi \sigma}},\non
(Q_m )^{\mu\nu} &=\frac{\pi K}{2\fues{1+\lambda}a^2} \, 
e^{\imath \fues{2\pi\sigma m+\sigma bk_2}}
D^{\mu\nu} \fues{\sqrt{\pi \sigma}}+c.c. ,
\end{align}
here the parameter $K$ is a measure of the strength of the coupling
between Landau bands 
\begin{eqnarray}\label{kkk1}
K=\frac{ma^2U_0\fues{1+\lambda}}{\hbar^2\pi}.
\end{eqnarray}
Taking into account these simplifications and that $D_N$ cancels out, we find
that  \Eq{har3} reduces to 
\begin{equation}\label{har5}
Q^- c_{m-1}+ Q_m c_m+Q^+ c_{m+1}=
\fues{\mca{E}+Ek_2+\sigma bEm}c_m.
\end{equation}

\begin{figure} [hbt]
\begin{center}
\includegraphics[width=3.0in]{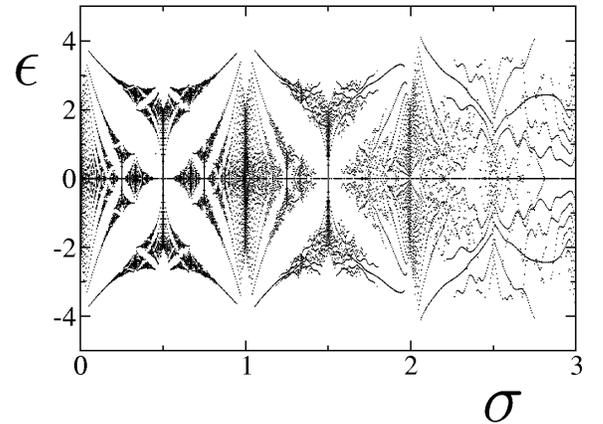}
\end{center}
\caption{The energy spectrum inside the lowest Landau level as a function
of the inverse magnetic flux $\sigma$. The energy  axis is rescaled to
$\epsilon={\cal E} /\left[U_0 e^{-\pi \sigma/2} \right]$. All values of
$k_1\in\yav{0,2\pi/qb}$  are included.
The parameters selected are:
$m^* = 0.067 m_e$, $a = 100 \, nm$, $U_0 = 0.5 meV$,
$E = 0.5 V/cm$, and $k_2 = \pi/2$. }
\label{figure1}
\end{figure}

\begin{figure} [hbt]
\begin{center}
\includegraphics[width=3.0in]{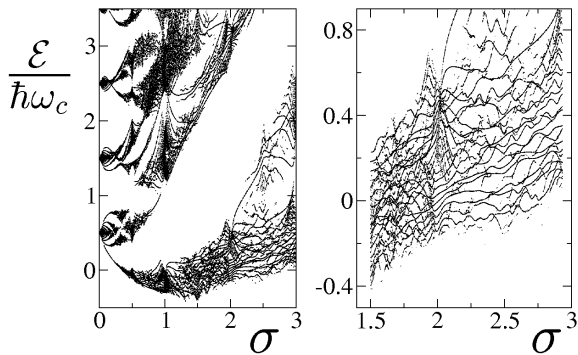} 
\end{center}
\caption{The energy spectrum for the four lowest Landau bands is plotted
as a function of  
magnetic flux $\sigma$. The  coupling strength $K =6$  yields important
Landau mixing. 
All values of $k_1\in\yav{0,2\pi/qb}$  are included. The parameters selected are:
$m^* = 0.067 m_e$, $a = 100 \, nm$, $U_0 = 1.2 meV$, $E = 0.5 V/cm$,
and $k_2 = \pi/2$. Seven  Landau levels are included in the calculation
in order to attain convergence.}
\label{figure2}
\end{figure}

\begin{figure} [hbt]
\begin{center}
\includegraphics[width=3.0in]{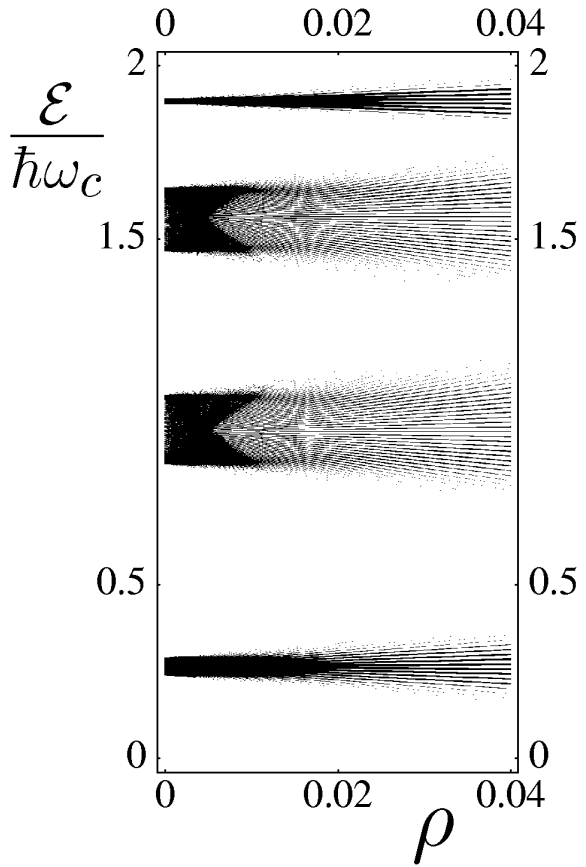}
\end{center}
\caption{Energy density plot  for the two  lowest Landau levels as a function
of the  electric field intensity ( $\rho = e a E/U_0$ ) for 
$\sigma=1/2$. All values of
$k_1\in\yav{0,2\pi/qb}$  are included. }
\label{figure3}
\end{figure}

%
\subsection{Numerical results}\label{numeric}
The generalized Harper equation is given by a tridiagonal infinite
recurrence relation \rep{har5}. If $\bs E$ is switched off, the equation becomes
periodic with period $p$, in this case the equation can be recast as a finite
$p L \times p L$ matrix that can be solved by a direct diagonalization.
However,  the introduction of $\bs E$ breaks the periodicity,  and most of the
methods used to solve the Harper equation
break down.     \Eq{har5} was solved  
using  an expansion of   the associated Green's
function into matrix continued fractions (MCFs) \cite{Risken1}. 
The energy spectrum is determined detecting the
change of sign in the Green's functions that appear in the vicinity
of a pole. The density of states can be obtained from
the expression 
\begin{equation}\label{gre5}
N\fues{\mca{E}}=
\frac{\imath}{\pi}\mathrm{Tr}\;
\tilde{G}\fues{\mca{E}}=
\mp\frac{1}{\pi}
\mathrm{Im}\yav{\mathrm{Tr}
\;G^{\pm}\fues{\mca{E}}} \, , 
\end{equation}
were the discontinuity in  Green's function is given by 
$\tilde{G}\fues{\mca{E}} =G^{+}\fues{\mca{E}} -G^{-}\fues{\mca{E}}$, 
and the retarded and advanced Green's functions are defined on the side limits 
$G^{\pm}\fues{\mca{E}} =\lim_{\epsilon \rightarrow 0}
G\fues{\mca{E} \pm \imath \epsilon}$. 
The numerical solution is obtained by truncating the iteration of the (MCFs) after the 
$M-$th term. 
The solution converges if $M$ is large  enough, in the calculations
we observe that in order to obtain a convergence with a precision
of one part in $10^7$, the cutoff can be selected  as
$M \approx 100$.

In our calculations we have used the effective mass $m^* = 0.067 m_e$ typical
for electrons in $GaAs$ and a superlattice
constant   $a = 100 \, nm$ \cite{Klit2}. The rescaled  energy spectrum
${\cal E} /\left[U_0 e^{-\pi \sigma/2} \right]$ is
shown in \Fig{figure1} for the lowest Landau level,
a weak modulation is considered ($U_0 = 0.5 meV$) so
the $\mu-\nu$ Landau mixing is negligible.
The electric field intensity is $E = 0.5 V/cm$ corresponding
to a ratio of the electric to periodic energy
$\rho = e a E/U_0 = 0.02$.   In the strong magnetic region
$\sigma\in\yav{0,1}$ the  Hofstadter butterfly is clearly depicted.
A distorted replica of the   butterfly spectrum can be still observed
in the region $\sigma\in\yav{1,2}$.
As the magnetic field loses intensity,  the effect of the electric field
becomes dominant, the butterfly is replaced by  discrete levels
separated in  regular intervals, i.e. a Stark ladder. 
Although, $\rho = e a E/U_0$ is small, there is a regime in which the
electric field dominates over the periodic contribution, we can trace
down the origin of this effect to  \Eqs{laguerres} and \rep{defQ}
showing that the periodic contribution is modulated by a factor
$e^{-\pi \sigma/2}$.
Consequently, in addition to a small Landau mixing, the condition
to preserve the butterfly spectrum  can be stated as
\begin{equation}\label{bound1}
e a E \ll U_0 e^{-\pi \sigma/2} \, .
\end{equation}
 This condition  restricts
the intensity of both the electric and magnetic fields,  these are estimated as
(inserting units):  $E \ll U_0/(e a) \sim 100 \, (U_0/meV) \, V/cm$  and
$B \geq \pi h /(e  a^2) \sim 0.6 T $. \Fig{figure2} shows the energy spectrum
as a function of $\sigma$ when a higher coupling strength ($K =6$)  produces
Landau mixing. For strong magnetic field the effect of the electric field is
small and the spectrum is very similar to that previously obtained by Petschel
and Geisel \cite{Geisel1}. Three regimes can be identified in the plot:
$(i)$ For strong magnetic field (small $\sigma$) the Landau bands are well
separated and butterfly structures are clearly identified.
$(ii)$ In the intermediate region ($\sigma  \sim 1$) the  periodic potential
induces Landau level overlapping,
$(iii)$ as $\sigma$ increase  very narrow levels appear as a result of the
electric field. The development of quasi discreet level induced by   the
electric field is  displayed   in \Fig{figure3}, here the energy spectrum as
a function of $E$ for the first two Landau levels is presented.
For $\sigma =1 /2$ every Landau level splits in two bands,
these bands evolve into  a series of quasi discreet levels as the
intensity of $E$ increases.

\begin{figure} [hbt]
\begin{center}
\includegraphics[width=3.0in]{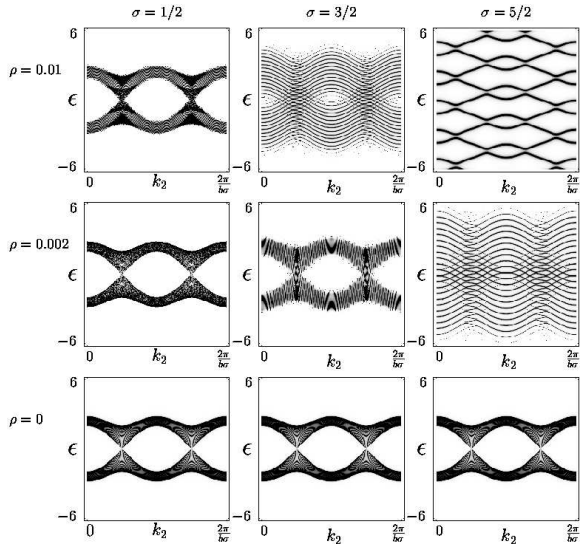} 
\end{center}
\caption{Energy density plot  for the two  lowest Landau levels as a function
of the  transversal pseudomomentum $k_2$. Three values of $\sigma=1/2,3/2,5/2$
are selected, so  two bands  appear when  $E = 0$. Three values of $E$
are considered: $\rho = e a E/U_0= 0, \,0.002,\, 0.01.$ }
\label{figure4}
\end{figure}

In \Fig{figure4} density plots of the density of states
as a function of the $k_2$ pseudomomentum and the quasienergy  ${\cal E}$
are presented. The density of states
was calculated from  Green's function discontinuity obtained from the
continued fraction expansion. 
Three values of $\sigma=1/2,3/2,5/2$ are selected, consequently   two bands
appear when  $\bs E = 0$. 
In the superior panels it is observed that both  as $E$ increases or $B$ 
decreases, the spectrum for a fixed value of $k_2$ evolves  into a set of
narrow bands. When  $k_2$ is varied the spectrum becomes almost continuous,
except for the small gaps that open between the bands. 
 
Based on the  previous results it follows  that the EMB model preserves the
band  structure. For weak  electric fields the bands are
grouped forming the `` butterfly  spectrum"; whereas as the intensity of $\bs E$
increases the bands are replaced by a  series of quasi discreet levels:
a ``magnetic a Stark ladder".

\subsection{Adiabatic approximation}\label{adiabatic}
In this section we exhibit an alternative expression for the effective \Sh
equation that governs the dynamics of the  system. Based on  this formalism
we shall  find an  approximated adiabatic solution, that  throws  further
insight into the physical results obtained in the previous  section. Let us
consider the generalized Harper equation \rep{har5},  taking into account
that the matrix $Q_m$ is periodic in  $m$ with period $p$, we find convenient
to define the unitary transformation 
\begin{align}\label{dia1}
d_l (\phi)  &= \sum_m \mb{U}_{l,m}(\phi)  \, c_m \, ,  \non
  \mb{U}_{l,m}\fues{\phi} & =\sqrt{\frac{q a}{2\pi}}
e^{\imath q a \phi  [ m/p ]} \, \delta_{l, m mod[p]}  \, ,
\end{align}
where $ [ m/p ]$ denotes the integer part of the number, and the delta enforces
the condition $ l = m, \, mod[p]$. Notice that whereas the index $m$ in $c_m$
runs over all the integers, the corresponding label of the state $d_l (\phi)$
takes the values $l =0,1,2, \dots ,p-1$. The new vector state $d_l (\phi)$
satisfies the periodicity condition 
\begin{equation}\label{periodi10}
d_l \left(\phi + {2 \pi \over q a} \right) \, = \, d_l (\phi) \,, 
\end{equation}
 and the transformation matrices  fulfill  the following properties 
\begin{align}
\sum_{m=-\infty}^\infty  \mb{U}_{l,m}\fues{\varphi}
\mb{U}_{m,l^\prime}^{\dag}\fues{\phi}&=
\delta_{l,l^\prime} \, \delt{\phi-\varphi},  \nonumber\\
\sum_{l=0}^{p-1} \int_{0}^{\frac{2\pi}{q a}}d\phi\;
\mb{U}_{m,l}^{\dag}\fues{\phi}\mb{U}_{l,m^\prime}\fues{\phi} &
=\delta_{m,m^\prime} .
\label{nor1}
\end{align}
Applying  this transformation to  the generalized  Harper equation
\rep{har5} yields
\begin{align}\label{hep10}
H_M &  \fues{k_1+\phi,  k_2} 
d\fues{k_1+\phi,k_2}  = \non
 & \fues{\mca{E}+Ek_2
-\imath E\frac{\partial}{\partial \phi}}
d\fues{k_1+\phi,k_2},
\end{align}
where the  Hamiltonean is   reduced to a   $p  \times p$  block  form 
\begin{equation}\label{hep16}
H_M\fues{k_1+\phi,k_2}=
\fues{\begin{array}{cccccccc}
  Q_0    &  Q^+   &     0     &   0       &  \dots  & Q^-  \\
 Q^-   &   Q_1    &   Q^+   &   0       &  \dots  &   0    \\
   0     &  Q^-   &   Q_2     &   Q^+   &  \dots  &   0    \\
 \vdots  &  \vdots  &  \vdots   &  \vdots   &  \ddots & \vdots \\ 
Q^+&    0     &     0     &    0      &  \dots  &Q_{p-1} \\
\end{array}}
\end{equation}
here  $Q^- \equiv Q^- \fues{k_1+\phi,k_2}$,
$Q_m\equiv Q_m\fues{k_1+\phi,k_2}$ and
$Q^+ \equiv Q^+ \fues{k_1+\phi,k_2}$.
Taking into account that each block in \rep{hep16} is a $L \times L$ matrix
determined by the Landau indexes,
we have obtained that  the transformation \rep{dia1}  reduces the infinite
dimensional representation of the \Sh equation to a $pL \times  pL$ representation.
The price is that   a non-local derivative term respect to  parameter $\phi$
has been added.  In the absence of electric field, \Eq{hep10}
represents a finite dimensional   eigenvalue problem, that can be solved by direct
diagonalization. We observe  from Eqs.  \rep{hep10} and \rep{hep16} that
$\phi$ appears to be directly  related the longitudinal quasimomentum  $k_1$,
in fact if we redefine $k_1+\phi\rightarrow k_1$  in \Eq{hep10} yields 
\begin{equation}\label{hep14}
H_M \fues{k_1,k_2}d\fues{k_1,k_2}=\fues{\mca{E}+Ek_2
-\imath E\frac{\partial}{\partial k_1}}d\fues{k_1,k_2}.
\end{equation}
This expression confirms the form of the  \Sh
equation  \rep{pro3} previously discussed, in which  the ``dynamics" is determined 
by a differential equation with respect 
to the longitudinal quasimomentum. However we now have an 
explicit  finite-dimensional matrix representation for the Hamiltonean.

Let the  ``instantaneous" eigenstates of $H_M$  be $ h^{(\alpha)}$ with
energies $\Delta^{(\alpha)}$, i.e.
\begin{equation}\label{ada0}
H_M   h^{(\alpha)} \left(k_1,k_2\right)=
\Delta^{(\alpha)}  \left(k_1,k_2\right)   h ^{(\alpha)} \left(k_1,k_2\right) \, , 
\end{equation}
where $\alpha$  labels  the band state of the non-perturbed problem.
A solution of  Eq. (\ref{hep14}) can  readily be obtained in the adiabatic
approximation as follows
\begin{align}
&  d^{(\alpha)} \left(k_1,k_2\right)=   h^{(\alpha)} \left(k_1,k_2\right) \times \non
& \exp\left[-\imath \frac{\cal E}{E} k_1
+\imath \frac{1}{E}\int_{0}^{k_1}d\phi   \Delta^{(\alpha)} \left(
\phi , k_2 \right)+\imath \gamma^{(\alpha)} \left(k_1,k_2\right)\right]  ,
\label{ada1}
\end{align}
where the Berry phase  $\gamma\left(\phi\right)$ is determined from the substitution 
of the previous  expression in Eq. (\ref{hep14})
\begin{equation}
\gamma^{(\alpha)} \left(k_1,k_2\right)= \,\imath  \int_0^{k_1}d\phi  \,
h^{(\alpha) \dagger} \left(\phi,k_2\right)    
\frac{\partial}{\partial \phi}  h^{(\alpha)}  \left(\phi,k_2\right) \, . 
\end{equation}
The energy eigenvalue is then determined by the periodicity
condition \rep{periodi10},  hence the change of the phase of the wave
function \rep{ada1} must be an integral multiple of $2\pi$, consequently
the  spectrum in the  adiabatic approximation is given as 
\begin{align} \label{enersp} 
\mca{E}^{(\alpha)} \fues{k_2}= & n E a \sigma
 + \frac{q a }{2\pi}\int_{0}^{2\pi/q a}dk_1 \Delta^{\alpha}
\fues{k_1,k_2} \non
& +\frac{q a E}{2\pi} \, \gamma^{(\alpha)}(2\pi/q a, k_2)
, \quad n=0,\pm 1, \dots
\end{align}

This result deserves some comments. The energy spectrum in the presence of
electric and magnetic field contains a series of discrete levels separated by 
\begin{equation}
\Delta \mca{E} =  E a \sigma = { h E \over a B},
\label{stark}
\end{equation}
where we have used \Eq{sig1} and restored units. These levels are similar to the Wannier levels that
appear when an electric field is applied to an electron in a periodic potential,
the energy separation being proportional to   $a E$.
In the present  case, the band structure parallel to the electric field is
replaced by a set of discrete steps, this ``magnetic-Stark"  ladder
are characterized by a separation proportional to the electric field intensity,
but inversely proportional to both the lattice separation  and the magnetic field. 
The  existence of these levels can be explained by the following argument.
In the presence of simultaneous $\bs E$ and $B$ fields,
the electron travels  between lattice points in a time $\tau = a B /E$.
As long as the electron does not tunnel into another band, the motion appears
as periodic, with frequency $\omega = 2\pi / \tau$,   corresponding to a series
of energy levels whose separation $\Delta \mca{E}  = \hbar \omega$ coincides with the
result in \Eq{stark}. This magnetic-Stark ladder is combined with the Hofstadter
spectrum represented by the second term in \Eq{enersp} and the contribution of
the Berry phase, the competition between these three factors was  discussed in
the numerical analysis of the previous  section.  We notice that  the Berry phase
is written as the integral of the longitudinal component of the Berry connection
$\mca{A}\sa\fuve{k}$ discussed by Kohmoto \cite{Khomo1} (see \Eq{pbe21}).

%
%
\section{Hall conductivity}\label{hall}
In order to calculate the electric current for a  system of
independent electrons let us consider the fermionic
field  \cite{Haken1}
\begin{equation}\label{efh3}
\psi\fuel{t}{x}= \sum\sa\int {d^2 k \over (2\pi)^2} \;
b\sa\fuve{k}\varphi\fusa{k}{t}{x},
\end{equation}
expanded in terms of the EMB states $ \varphi\sa$, that according to
the Floquet \Eq{flo4}  are related with  the solution of the \Sh as
$\Psi\sa = e^{- \imath {\cal E} t} \varphi\sa$.
The  creation and  annihilation operators  $b^\dag\sa\fuve{k}$ and 
$b\sa\fuve{k}$  fulfill the usual fermion anti-commutation relations
$\llav{b\sa\fuve{k},b^\dag\sap\fuve{k\pr}}=
\delta_{\alpha,\alpha\pr}\deltus \, ,$ and zero for
other anticommutators. The current density operator 
$\psi^\dag \bs{\hat{J}} \psi$ takes the form 
\begin{equation}\label{cor0}
\bs{\hat{J}}= 
\sum_{\alpha,\alpha\pr}\int {d^2k\pr \over (2\pi)^2} \int {d^2k \over  (2\pi)^2} \;
b^\dag\sa\fuve{k\pr}b\sa\fuve{k}
\braket{\alpha\pr,\bs{k}\pr}{\bs{\Pi}}{\alpha,\bs{k}},
\end{equation}
where
\begin{equation}
\braket{\alpha\pr,\bs{k}\pr}{\bs{\Pi}}{\alpha,\bs{k}}
=\int d^2x\;\varphi^{*}\fuap{k}{t}{x}
\bs{\Pi}\varphi\fusa{k}{t}{x}.
\end{equation}
In order to evaluate the previous expression, we first consider   the matrix  element
of $\bs x$. Using the replacement $\varphi\fuen{k}{t}{x}=
e^{\imath \bs{k}\cdot \bs{x}}
u\fuen{k}{t}{x},$ that follows from the Bloch theorem \Eq{flo4}, it can be readily
demonstrated that 
\begin{align}\label{pos21}
\braket{\alpha\pr,\bs{k}\pr}{\bs{x}}{\alpha,\bs{k}}
=  -\imath \nas{k} &
\deltus \delta_{\alpha\pr,\alpha}  \non
& +  \deltus  \bscal{A}^{\alpha\pr,\alpha}\fuve{k},
\end{align}
with the definition 
\begin{equation}\label{pbe21}
\bscal{A}^{\alpha\pr,\alpha}\fuve{k}=  \frac{(2 \pi)^2}{q a^2}
\int_{\rm MUC } d^2x\; u^{*}\fuap{k}{t}{x}
\imath \nas{k}u\fusa{k}{t}{x}.
\end{equation}
The subscript means that the spatial integration is restricted to a magnetic unit
cell.  We observe that the diagonal element 
$ \bscal{A}\sa\fuve{k}=\bs{\mca{A}}^{\alpha,\alpha}\fuve{k}$ is the induced Berry
connection \cite{Khomo1,Simon1,Avron1}.
The derivation of  \Eq{pos21} for Bloch 
states is well known (see for example\cite{Callaway}),  the demonstration is easily
extended for the  MB states \cite{MingChe1} and also for the EMB states \cite{Kunold2}.
Using the relation
$\bs{\Pi}=\imath \yav{H,\bs{x}}$, the effective \Sh equation \rep{pro3} and
the result in \rep{pbe21} one finds 
\begin{align}\label{pbe3}
& \braket{\alpha\pr,\bs{k}\pr}{\bs{\Pi}}{\alpha,\bs{k}} = \non
 &\yav{\bs{\Pi}\sa\fuve{k}
\delta_{\alpha\pr,\alpha}+\bs{\Pi}_{\alpha\pr,\alpha}\fuve{k}
\fues{1-\delta_{\alpha\pr,\alpha}}}\deltus,
\end{align}
where
\begin{align}
\bs{\Pi}\sa\fuve{k}&=
\nas{k} \mca{E}\sa\fuve{k}
-\bs{E}\cdot\nas{k}\bscal{A}\sa\fuve{k}, \num \\
\bs{\Pi}_{\alpha\pr,\alpha}\fuve{k}&=
\imath \yav{\mca{E}\sap\fuve{k}-\mca{E}\sa\fuve{k}+\imath \bs{E}\cdot\nas{k}}
\bscal{A}^{\alpha\pr,\alpha}\fuve{k}.
\end{align}
Substituting  \rep{pbe3} into
\rep{cor0} the electric current becomes
\begin{align}\label{cor1}
& \bs{\hat{J}}= \sum\sa\int   {d^2k \over  (2\pi)^2}\;\;
b^\dag\sa\fuve{k} b\sa\fuve{k} 
\yav{\nas{k}\mca{E}\sa\fuve{k}
-\bs{E}\cdot\nas{k}\bscal{A}\sa\fuve{k}} \non
&+ \imath \sum_{\alpha,\alpha\pr}\int {d^2k \over  (2\pi)^2} \;\;
b^\dag\sa\fuve{k} b\sap\fuve{k}
\yav{\mca{E}\sap\fuve{k}-
\mca{E}\sa\fuve{k}}\bs{\mca{A}}^{\alpha\pr,\alpha}\fuve{k}.
\end{align}
At  very low temperature the 
electrons occupy all the  levels that are below  the Fermi energy.
So the  state of the system in the independent particle approximation 
can be  written in the form 
\begin{equation}\label{nml1}
\ket{\Psi}= \prod_{\alpha\le \nu_F}
\prod_{\bs{k}} b^\dag\sa  \fuve{k}\ket{0},
\end{equation}
where  $\ket{0}$ is the vacuum,  $\nu_F$ is 
the number of filled bands below the Fermi level and  the state is normalized
in such a way that the number of charge carriers is given as
$\mca{N} = \nu_F \, S /(q b^2)$, where $S$ is the area of the sample. 
For the state \rep{nml1}   the electric  current  becomes
\begin{equation}\label{tom1}
\bs{J}=  \sum_{\alpha\le \nu_F}
\int {d^2k \over  (2\pi)^2} \;
\yav{\nas{k}\mca{E}\sa\fuve{k}
-\bs{E}\cdot\nas{k}\bscal{A}\sa\fuve{k}}.
\end{equation}
The first term on the right hand side represents the usual velocity
group contribution. Additionally there is a novel contribution arising
from gradient of the Berry connection along the longitudinal electric
field direction.

In \ref{adiabatic}  we discussed the solution of the effective \Sh
\rep{hep14} equation in terms of the adiabatic approximation. However, if we
consider the matrix element of \Eq{hep14} taken between the exact  EMB  states
that solve \Eq{hep14}, we find that the energy eigenvalue  fulfills the relation 
\begin{equation}\label{enp1}
\mca{E}\sa\fuve{k}=\Delta\sa\fuve{k}
+\bs{E}\cdot\bscal{A}\sa\fues{k_1,k_2}-Ek_2,
\end{equation}
where 
\begin{equation}\label{enp2}
\Delta\sa\fuve{k}=
\langle u_{\alpha,\bs{k}} \vert   H_M \vert u_{\alpha,\bs{k}} \rangle .
\end{equation}
The  eigenvalues in  \rep{enp1} and the previous  results in
\rep{enersp} look similar; however the adiabatic approximation
makes use of the approximate instantaneous  solution 
\rep{ada1} in order to evaluate the quantities that appear
in \Eq{enersp}. Instead   the evaluation of 
\rep{enp1}  requires the exact solution of the \Sh equation
\rep{hep14}. 
Substituting \Eq{enp1} into the expression \rep{tom1} for the
current, this reduces to the form 
\begin{align}\label{xyz1}
\bs{J}=  & \sum_{\alpha\le \nu_F} 
\int {d^2k \over (2 \pi)^2 } \; \left(
\nas{k}\yav{\bs{E}\cdot \bscal{A}\sa\fuve{k} -Ek_2}
- \bs{E}\cdot\nas{k}\bscal{A}\sa\fuve{k} \right)  \non
= & {\nu_F \over 2 \pi p} E\bs{e}_T-\frac{1}{2 \pi p }\bs{E}\times 
\sum_{\alpha\le \nu_F} \int \frac{d^2k}{2\pi}\;
\bs{\Omega}\sa\fuve{k} \, , 
\end{align}
the Berry curvature $\bs{\Omega} \sa = \bs{e}_3 \Omega \sa$  is related to  the 
potential vector $\bscal{A}\sa$ by 
\begin{equation}\label{spn1}
\Omega \sa\fuve{k} 
=\yav{\nas{k}\times\bscal{A}\sa\fuve{k}}_3 \, . 
\end{equation}
\Eq{xyz1} shows that the longitudinal magnetoresistance 
exactly cancels, regardless of the value of $\bs E$. On the other hand, 
the Hall conductance  can be read  from  \Eq{xyz1}  and  written (restoring units)
in a compact form as 
\begin{equation}\label{cnd4}
\sigma_H=    \frac{e^2}{h} \sum_{\alpha\le\nu_F}  \sigma\sa \, , \qquad \qquad 
\sigma\sa   = \frac{1}{ p} \left( 1 - q \, \eta\sa  \right) \, ,  
\end{equation}
where $\eta\sa$ is the $\alpha$ subband contribution to the conductance given as
\begin{equation}\label{www5}
\eta\sa = {1 \over q} \int\frac{d^2k}{2\pi}
\yav{\Omega\sa\fuve{k}}_3 ={ 1 \over q} \oint_{\rm CMBZ}
\frac{\bs dk}{2\pi}\cdot
\bscal{A}\sa\fuve{k} \, , 
\end{equation}
here CMBZ denotes the contour of the MBZ and
the Stoke's theorem was used to write  the second equality.
We find it convenient to multiply and divide by $q$ the
contribution of $\eta\sa$ that appears in \Eq{cnd4},
the reason is that the Brillouin zone $k_1\in\yav{0,2\pi/qa}$,
$k_2\in\yav{0,2\pi/a}$: can be divided in $q$ restricted Brillouin 
zones, RMBZ: $k_1\in\yav{0,2\pi/qa}$,
$k_2\in\yav{2\pi j/qa,2\pi (j + 1)/qa}$ with $j = 0,...., q-1$.
The wave function satisfies periodicity conditions that connect
the solutions in different RMBZ and thus the integrals in
\Eq{www5} can be evaluated in any of the RMBZ giving
the same result \cite{Kunold2}.

\begin{figure} [hbt]
\begin{center}
\includegraphics[width=3.0in]{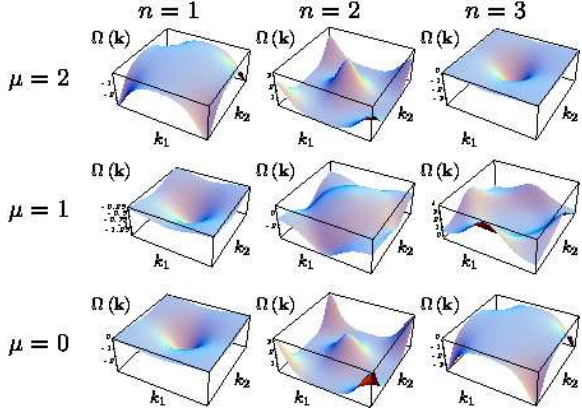}
\end{center}
\caption{Berry  curvature distribution on the MBZ  for the
first three Landau levels. 
$\alpha$ labels the Landau level $\mu$ and the corresponding subband $n$.
The inverse magnetic flux is $\sigma=1/3$, so there are three
subbands for each Landau level. The  coupling strength is  $K =0.2$. }
\label{figure5}
\end{figure}

An expression  similar to Eq. (\ref{xyz1}) was originally discussed
by Thouless \etal \cite{Thou1} and Kohmoto \cite{Khomo1}.
However, their work is based on the Kubo linear response theory,
consequently only terms linear in  $E$ appear in that case.
Our result reduces to that of Thouless and
Kohmoto if we drop the electric  field in \Eq{hep14},
in other words  if we evaluate the conductivity in \Eq{www5}
using the  zero order solution  \Eq{ada0}.
As far as the gap condition is satisfied the argument that shows that the expression in 
\Eq{www5} is related to the first Chern number can be applied to prove that $\eta\sa$ is quantized.
However as the electric field is increased the high density of levels makes it practically impossible 
to resolve the conductance contributions for each miniband, as far as the original Hofstadter miniband 
is concerned this fact can be evaluated as a nonlinear contribution to  $\sigma\sa$ that will lead to a break down 
of its quantization.
Higher order  corrections to the Hall conductivity can be
calculated by a perturbative evaluation of \Eq{hep14}.
We observe that the  term  $E( k_2-i\partial/\partial k_1)$
in \Eq{hep14} can be considered as a perturbative potential;
hence a series expansion in $E$ can be worked out. 
The Hall conductance for the $\alpha-$band is expanded as 
\begin{equation}\label{excon}
\sigma\sa =  \sigma\sa^{(0)}  +  \sigma\sa^{(1)}  E  +   \sigma\sa^{(2)}
 E^2 \, + \,  \dots\, . 
\end{equation} 

\begin{figure} [hbt]
\begin{center}
\includegraphics[width=3.0in]{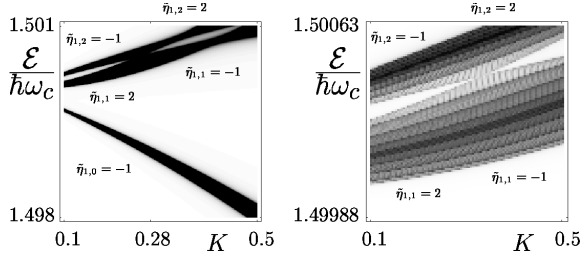}
\end{center}
\caption{Energy density plot as a function the coupling strength $K$ for the
three minibands of the $\mu=1$ Landau level. $\sigma =1/3$. Notice that at
$K=0.28$  the two upper bands crosses and the corresponding 
values of $\eta\sa^{(0)}$ are exchanged. }
\label{figure6}
\end{figure}

\begin{figure} [hbt]
\begin{center}
\includegraphics[width=3.0in]{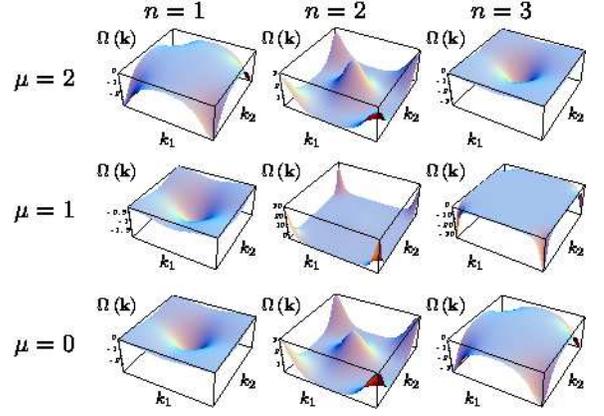} 
\end{center}
\caption{Berry  curvature distribution on the MBZ  for the
first three Landau levels.  $\alpha$ labels the Landau level and the
corresponding subband. $\sigma=1/3$, so again  there are three subbands
for each Landau level but now $K =0.28$ .  Notice that the curvature
distribution for the  bands of the first Landau level are modified as compared
with those in \Fig{figure5}. The reason  is  that   $K$ lies to the right of the 
point  (see \Fig{figure6}) at which the   second and third minibands crosses.}
\label{figure7}
\end{figure}

Starting with the zero-order solution  of \Eq{hep14}
$u\sa^{\fues{0}}\fues{t,\bs x}=\langle t, \bs x \vert \alpha^{(0)} \rangle$,
the leading contribution to the Berry connection is simply given by 
\begin{equation}\label{fisrto}
{\bscal A}\sa^{(0)} =\imath
\brakete{\alpha^{(0)}}{\nabla_{\bs k}}{ \alpha^{(0)} }.
\end{equation} 
Then based on definition \rep{www5} and the condition
\rep{flo6} fulfilled by the wave function on the boundary
of the MBZ, it follows that 
$\eta\sa^{(0)} $ is the change of the wavefunction's phase around
the integration loop. The wave function has zeros in the interior
of the MBZ thus $\eta\sa^{(0)} $ must be an integer.
$\eta\sa^{(0)} $  is evaluated numerically.
The contour  integration 
around the MBZ  in \Eq{www5}   is numerically unstable;  thus it is more
convenient to evaluate the surface integral of  the Berry curvature.
>From \Eqs{fisrto} and  \rep{spn1} $\Omega\sa^{(0)}$ is written as 
$ \Omega\sa^{(0)}  =\imath \left [
\langle \partial_{k_1} \alpha^{(0)} \vert  \partial_{k_2} \alpha^{(0)}
\rangle -  \langle \partial_{k_2} \alpha^{(0)} \vert  \partial_{k_1}
\alpha^{(0)} \rangle \right]$,
inserting a complete state
$\sum_{\beta} \vert \beta^{(0)}  \rangle \langle \beta^{(0)} \vert$  inside
the dot product and using the  identity 
\begin{equation}\label{idenaux}
\left(  {\cal E}_{\beta}^{(0)} -{\cal E}_{\alpha}^{(0)} \right)
\left\langle \alpha^{(0)}  \vert {\partial \over \partial_{k_i} }
\beta^{(0)} \right\rangle = 
\braket{\alpha^{(0)}  }{{\partial H \over \partial k_x }}{ \beta^{ (0)}},
\end{equation} 
that can be deduced from the partial derivative of  \Eq{ada0},
the Berry curvature can be written in the form 
\begin{align}\label{eqh8}
& \left[ \Omega\sa^{(0)}  \fuve{k} \right] = \non
& \imath  \,   \sum_{\beta \ne\alpha}
\left[{\braket{\alpha^{(0)}  }{{\partial H \over \partial k_x }}{ \beta^{ (0)}}
\braket{\beta^{(0)}}{{\partial H \over \partial k_y }}{\alpha^{(0)}} \,
\over  \left[ \omega_{ \beta \alpha}  \right]^2 } - \, {\rm C. c.} \right],
\end{align}
where $\omega_{ \beta \alpha} =  {\cal E}_{\beta}^{(0)} -{\cal E}_{\alpha}^{(0)}$. 
 \Eq{eqh8}   is well suited for numerical evaluations. 
\Fig{figure5} shows the distribution of  the Berry curvature plotted in the magnetic
Brillouin zone when the  inverse magnetic flux is
selected as $\sigma=1/3$ and the coupling strength $K =0.2$.
In this case, each Landau level  $\mu$ is splitted  in three subbands, results
are presented  for  the first three Landau levels. It is  verified that
$\eta\sa^{(0)}$ always yields an integer value; here $\alpha = (\mu, n)$ labels
the Landau level $\mu$ and the $n$ internal miniband.  In this example we find:
$\eta_{\mu,1}^{(0)}=1$, $\eta_{\mu,2}^{(0)}=-2$, and $\eta_{\mu,3}^{(0)}=1$,
regardless of $\mu$.  The numbers $q$ y $p$ are relatively prime, so there must
be integer numbers $u$ y $v$ such that  $up+vq=1$.  We can identify
$v\equiv \eta \sa^{(0)}$,  hence   $(1-q\eta\sa^{(0)})/p$
is also a integer, but according to  \Eq{cnd4}  this number is  the Hall
conductance of the subband $\alpha$; thus $\sigma\sa^{(0)}$ is quantized in
units of $e^2/h$. As the coupling strength $K$ increases some of the bands
crosses (see \Fig{figure6}) and the values of  $ \eta \sa^{(0)}$
are exchanged. \Fig{figure7} shows the plot of $\Omega\sa^{(0)}$ for 
$\sigma=1/3$ and  $K =0.28$. The Berry curvatures for the first Landau level
are modified, now: $\eta_{1,1}^{(0)}=1$, $\eta _{1,2}^{(0)}=1$,  and
$\eta_{1,3}^{(0)}=- 2$. The profiles of $\Omega\sa^{(0)}$ for small values of $K$ coincide 
with those that appear in the semiclassical dynamics approach introduced by Chang and Niu \cite{MingChe1}, our 
results include the effects produced by Landau mixing.

Table~\ref{tabla1} shows the values of $\eta\sa^{(0)}$
 for various  selections of  $\sigma$ when $K$ is small.
If the coupling between Landau levels is small, the following sum rules
are satisfied
\begin{equation}\label{sumrule1}
\sum_n \eta_{\mu,n}^{(0)} =0, \qquad \qquad \sum_n \sigma^{(0)}_{\mu,n}=1.
\end{equation}
here  the sum includes all the subbands  within a given Landau level.
The last  result guarantees  that the Hall conductivity of  a completely
filled Landau level takes the value $e^2/h$. 
However as the Fermi energy sweeps through a Landau level, the
partial contributions of the subband, although integer 
multiples of $e^2/h$, do not follow a monotonous behavior.
If the Fermi energy lies in the  $n$-minigap, the accumulated
conductance $\zeta\sa$ is defined as
$\zeta_{\mu,n}=\sum_{j =1}^{j=n}\sigma_{\mu,j}.$ The $n$-minigap conductance
satisfies the Diophantine equation $ n = \lambda \, q + p \, \zeta_{\mu,n} $,
$(\vert \lambda\vert \le p/2)$ in agreement with the results obtained by
Thouless \etal \cite{Thou1}.

We now turn our attention to the higher order correction to the Hall conductance.
>From the first order corrections to the wave function, the corresponding contribution 
to the conductance is worked out as 
\begin{align}\label{correction1}
& \eta\sa^{(1)}  = \non &  \sum_{\beta \ne\alpha} \oint_{\rm CMBZ} \frac{\bs dk}{2\pi}\cdot
\left[{  \langle \alpha^{(0)} \,
\vert {\partial H \over \partial { \bs k} } \vert  \, \beta^{(0)}   \rangle
\langle \beta^{(0)}  \, \vert {\partial H \over \partial k_y } \vert \,
\alpha^{(0)}  \rangle \over \left[ \omega_{\alpha \beta}  \right]^3  }
\, + \, C. c. \right].
\end{align}
As previously  discussed, the wave function evaluated at opposite
edges of the MBZ differs at most by a phase function $f_i\fuve{k}$
(see \Eq{flo6}), this phase exactly cancels in \Eq{correction1},
hence  $\sigma\sa^{(1)} = 0$ and there is no ${\cal O} (E^2)$
correction to the Hall current. 

The evaluation of $\eta\sa^{(2)}$ requires the first and second order
wave function contributions, then it leads in principle to a lengthly
expression, however the result simplifies if we notice that any term
in which the wave function cancels will not contribute, because the
corresponding contour integral around the MBZ will vanish.
Then including only the   relevant contributions the second order
correction to  the Berry connection can be worked out as   
\begin{align}\label{correction2}
 \left[ {\cal A }\sa^{(2)} \right]_i  & =   \sum_{\beta \ne\alpha}
\left(  {\cal A }\sa^{(0)} -  {\cal A }_\beta^{(0)} \right)_1 \times  \non &  
 \left[{ \langle \alpha^{(0)} \,
\vert {\partial H \over \partial { k_1} } \vert  \, \beta^{(0)}   \rangle
\langle \beta^{(0)}  \, \vert {\partial H \over \partial k_i } \vert \,
\alpha^{(0)}  \rangle  \, \over \left[ \omega_{\alpha \beta}  \right]^4  }
+  \,\left(1 \leftrightarrow i \right)  \right].
\end{align}

In the numerical evaluation of $\eta_\alpha^{(2)}$  we find convenient to
scale the energy as $\epsilon ={\cal E} /\left[U_0 e^{-\pi \sigma/2} \right]$,
then as discussed in \ref{numeric} the spectrum for a given Landau band
(for $\bs E = 0$) lies in the region $ \epsilon \in (-4,4)$.
In this case  $\eta_\alpha^{(2)}$ takes the form
$\eta_\alpha^{(2)} = 2\pi  \left( e l_0 / U_0 \right)^2 e^{\pi \sigma }%
{\tilde \eta}_\alpha^{(2)}$, where ${\tilde \eta}_\alpha^{(2)}$
is dimensionless. The contour  integration of \rep{correction2}  around
the MBZ yields
the second order correction  to the Hall conductance.
The expectation values of the partial derivatives of the Hamiltonean
are numerically evaluated in a similar fashion as those terms
in \Eq{eqh8}.  As already mentioned  the relevant part
of $ {\bscal A }\sa^{(0)}$ is the phase, that integrated around
the MBZ yields the leading order contribution 
to $\sigma_H$. Although  the numerical evaluation of this
phase is unstable,  it was previously explained how to determine
its integral around the MBZ, some number are quoted
in Table~\ref{tabla1}. We can now exploit a particular
gauge invariance of the problem. Suppose that $u\sak$
satisfies the \Sh  \rep{hep14}, then so does 
$u\sak = e^{\imath g\fuve{k}} u\sak$, where $g\fuve{k}$ is a smooth
function of $k_1$ and $k_2$. A  valid    gauge transformation, must not be
singular, in such a way that its  topological  signature ($\eta^{(0)}$)
remains untouched.
Then, along a contour  $\Gamma$, the phase of  $u\sak$
can be choose in such a way that the Berry connection
takes the form
${\bscal A }\sa^{(0)}=2\pi\eta\sa^{(0)}{\bs h}\fuve{k}+\imath%
\nabla_{\bs k} \rho \fuve{k}$, where $ {\bs h}\fuve{k}$
is a selected  function of $\bs k$  which integrates
around the contour of the MBZ to 1.
The term $\nabla_{\bs k} \rho\fuve{k}$ does not modify the value of
$\eta\sa^{(0)}$, in particular it can be adjusted to have the same value
in opposite sides of  the MBZ. The gauge transformation cannot be
implemented in all the magnetic Brillouin zone, it will be valid
only if $\Gamma$ encloses all the phase wave function singularities
(zeroes of $u_\alpha$); we are interested in the case in which
$\Gamma $ coincides with  CMBZ. With all these considerations
the second order contribution to the Hall conductance can be recast as 
\begin{align}\label{correction22}
& {\tilde \eta}\sa^{(2)}  = \sum_{\beta \ne\alpha} \oint_{\rm CMBZ}  d\bs{k} \cdot
\left( \eta\sa^{(0)} -  \eta_\beta^{(0)} \right) h_1\fuve{k} \times \non &
\left[{ \langle \alpha^{(0)} \,
\vert {\partial H \over \partial { k_1} } \vert  \, \beta^{(0)}   \rangle
\langle \beta^{(0)}  \, \vert {\partial H \over \partial {\bs k} } \vert \,
\alpha^{(0)}  \rangle \over \left[ \omega_{\alpha \beta}  \right]^4  } \,
+  \,\left(1 \leftrightarrow i \right)  \right].
\end{align}
A suitable  function ${\bs h}\fuve{k}$ can be selected as the
gauge potential of a unit flux tube in $k-$space:
${\bs h}\fuve{k} = {1 \over 2 \pi} (k_1, - k_2)/(k_1^2 + k_2^2)$.
In Table~\ref{tabla1} we quote some of the values obtained 
for ${\tilde \eta}\sa^{(2)}$ for the selected Landau subbands.
It is observed that as the internal structure of band increase,
the corrections to the conductance of the internal minibands
increases abruptly.
It can be easily proved from \Eqs{correction2} and \rep{correction22}
that the following sum rule applies 
\begin{equation}\label{sumrule2}
\sum_n \eta_{\mu,n} ^{(2)} =0.
\end{equation}
Consequently, even if each miniband presents
$\sigma_H^{(2)} \propto e^3 /U_0^2 \, B$ corrections 
to the Hall conductance, the contributions from a completely filled Landau level
cancels exactly,  and the Hall resistance 
quantization remains valid to order ${\cal O} (E^3)$. The nonlinear correction to
$\sigma_H$ is then expected to be more important for the minibands. A critical
electric field $E_c$ can be 
defined  as the the value at which the second order contribution
$\eta\sa^{(2)}$ becomes a fraction $f$ 
of  $\eta\sa^{(0)}$, it is expected that the quantization of the  Hall
conductivity  breaks down if the field 
becomes larger that $E_c$. According  to \Eq{excon} $E_c$ is estimated as 
\begin{equation}\label{ebd}
E_c \approx \,  \sqrt{f \,  {\eta^{(0)} \over  \eta^{(2)} }}  =  \, 
\,  \frac{U_0}{e \, l_0} \,e^{-\pi \sigma/2} \sqrt{ f \, { \eta^{(0)}
\over  \tilde \eta^{(2)} }}.
\end{equation}
Using the results in Table~\ref{tabla1} for the central miniband, and assuming
a required quantization precision of $f \sim 10^{-2}$,
$a =100 nm$ and $U_0= 0.5 meV$, the critical electric field   is  estimated
as $E_c \sim 215\,  V/cm$ for $\sigma = 1/3$, 
$E_c \sim 45 \,  V/cm$ for $\sigma = 1/5$ and $E_c \sim 23\,  V/cm$
for $\sigma = 1/7$. Thus,  the nonlinear effects are enhanced
as the number of internal minibands increases. 

\begin{table}
\begin{center}
\label{tabla1}
\begin{tabular}{c@{$\,\,\,\,$}c@{$\,\,\,\,$}r@{$\,\,\,\,$}|
      @{$\,\,\,\,$}c@{$\,\,\,\,$}c@{$\,\,\,\,$}r@{$\,\,\,\,$}|
      @{$\,\,\,\,$}c@{$\,\,\,\,$}c@{$\,\,\,\,$}r}
\hline
\multicolumn{3}{c}{$\sigma=1/3$}&
\multicolumn{3}{c}{$\sigma=1/5$}&
\multicolumn{3}{c}{$\sigma=1/7$}\\
\hline
\multicolumn{3}{c}{$q=1$, $p=3$}&
\multicolumn{3}{c}{$q=1$, $p=5$}&
\multicolumn{3}{c}{$q=1$, $p=7$}\\
\hline\hline
 $n$ & ${\eta}^{\left(0\right)}\sa$ & $\tilde{\eta}^{\left(2\right)}\sa$ &
 $n$ & ${\eta}^{\left(0\right)}\sa$ & $\tilde{\eta}^{\left(2\right)}\sa$ &
 $n$ & ${\eta}^{\left(0\right)}\sa$ & $\tilde{\eta}^{\left(2\right)}\sa$\\
\hline
$3$ &  $1$ &  $0.416$ & $5$ &  $1$ &  $-0.018$ & $7$ &  $1$ &    $0.0239$ \\
\hline
    &      &          &     &      &           & $6$ &  $1$ &    $0.2760$ \\
\cline{7-9}
    &      &          & $4$ &  $1$ &  $-4.152$ & $5$ &  $1$ & $-110.2766$ \\
\cline{4-9}
$2$ & $-2$ & $-0.328$ & $3$ & $-4$ &  $17.836$ & $4$ & $-6$ &  $117.3056$ \\
\cline{4-9}
    &      &          & $2$ &  $1$ & $-13.660$ & $3$ &  $1$ &   $-7.3303$ \\
\cline{7-9}
    &      &          &     &      &           & $2$ &  $1$ &    $0.0011$ \\
\hline
$1$ &  $1$ & $-0.087$ & $1$ &  $1$ &  $-0.006$ & $1$ &  $1$ &    $0.0003$ \\
\hline

\end{tabular}
\caption{Zero and second order coefficients $\eta\sa^{(0)}$, and $\tilde{\eta}\sa^{(2)}$  ($\alpha=(\mu,n)$)
for $\sigma=1/3$, $1/5$ and $1/7$.   The miniband conductivity  $\sigma\sa$  is obtained from \Eq{cnd4}.
The coupling between Landau levels is small, so the results are the same, regardless  of the $\mu$-Landau level.
The sum rules \rep{sumrule1} and \rep{sumrule2}  are easily verified. 
The   coefficients are symmetric respect to $\sigma=1/2$,  the values for  $\sigma \to 1 - \sigma$ 
are obtained with the replacement $\eta\sa^{(0)} \to - \eta\sa^{(0)}$
and $\tilde{\eta}\sa^{(2)} \to \tilde{\eta}\sa^{(2)}$ respectively ; however  $\sigma\sa$  does not share this 
symmetry .}
\end{center}
\end{table}

\section{Summary} 
In conclusion, we address the  electric-magnetic problem, and the
calculation of the Hall conductance beyond the linear response approximation. 
We presented a thoroughly discussion of the  symmetries of the EMB problems
and of the construction of the 
wave function and effective secular equation. The dynamics of the system is
governed  by   a finite difference  equation that exactly includes the  effects of:
an arbitrary periodic potential,  an electric field orientated in a commensurable
direction of the lattice,  and  coupling between Landau levels.   In addition to
the  broadening of the Landau levels induced by the  periodic potential,
the effect of the  the electric field in the energy spectrum is to superimpose
equally spaced discrete levels; in this ``magnetic Stark ladder" the energy
separation is an integer multiple of  $ h E / a B $. A closed expression
for the Hall conductance, valid to all orders in $\bs E$ is obtained. 
The leading order contribution is quantized  in units of $e^2/h$.
The first order correction  exactly vanishes, while  the  second order
correction shows  a  $\sigma_H^{(2)} \propto e^3 /U_0^2 \, B$
dependence. From the  sum rule \rep{sumrule2} it follows  that $\sigma_H^{(2)}$
cancels for a completely filled Landau band. Hence the nonlinear correction to
the conductance is expected to 
be of order  ${\cal O} (E^2)$  for a filled miniband, whereas for a complete
Landau level the correction is expected to be of order
${\cal O} (E^3)$.  We find that in  order to preserve the self-similarity
structure of the Butterfly spectrum,
the electric field is restricted according to \Eq{bound1},
 and additional restriction arises
from the condition that the nonlinear correction to the Hall conductance
remains small as  given by \Eq{ebd}.
 
\acknowledgments We acknowledge the partial financial support endowed by
CONACyT through grants No. \texttt{G32736-E}, \texttt{U42046} and
\texttt{42026-F}.

\bibliography{article}

\begin{thebibliography}{32}
\expandafter\ifx\csname natexlab\endcsname\relax\def\natexlab#1{#1}\fi
\expandafter\ifx\csname bibnamefont\endcsname\relax
  \def\bibnamefont#1{#1}\fi
\expandafter\ifx\csname bibfnamefont\endcsname\relax
  \def\bibfnamefont#1{#1}\fi
\expandafter\ifx\csname citenamefont\endcsname\relax
  \def\citenamefont#1{#1}\fi
\expandafter\ifx\csname url\endcsname\relax
  \def\url#1{\texttt{#1}}\fi
\expandafter\ifx\csname urlprefix\endcsname\relax\def\urlprefix{URL }\fi
\providecommand{\bibinfo}[2]{#2}
\providecommand{\eprint}[2][]{\url{#2}}

\bibitem[{\citenamefont{Klitzing et~al.}(1980)\citenamefont{Klitzing, Dorda,
  and Pepper}}]{Klit1}
\bibinfo{author}{\bibfnamefont{K.}~\bibnamefont{Klitzing}},
  \bibinfo{author}{\bibfnamefont{G.}~\bibnamefont{Dorda}}, \bibnamefont{and}
  \bibinfo{author}{\bibfnamefont{M.}~\bibnamefont{Pepper}},
  \bibinfo{journal}{Phys. Rev. Lett.} \textbf{\bibinfo{volume}{45}},
  \bibinfo{pages}{494} (\bibinfo{year}{1980}).

\bibitem[{\citenamefont{Tsui et~al.}(1982)\citenamefont{Tsui, St{\"o}rmer, and
  Gossard}}]{Tsui1}
\bibinfo{author}{\bibfnamefont{D.~C.} \bibnamefont{Tsui}},
  \bibinfo{author}{\bibfnamefont{H.~L.} \bibnamefont{St{\"o}rmer}},
  \bibnamefont{and} \bibinfo{author}{\bibfnamefont{A.~C.}
  \bibnamefont{Gossard}}, \bibinfo{journal}{Phys. Rev. Lett.}
  \textbf{\bibinfo{volume}{48}}, \bibinfo{pages}{1559} (\bibinfo{year}{1982}).

\bibitem[{\citenamefont{Thouless et~al.}(1982)\citenamefont{Thouless, Kohmoto,
  Nightingale, and den Nijs}}]{Thou1}
\bibinfo{author}{\bibfnamefont{D.}~\bibnamefont{Thouless}},
  \bibinfo{author}{\bibfnamefont{M.}~\bibnamefont{Kohmoto}},
  \bibinfo{author}{\bibfnamefont{M.~P.} \bibnamefont{Nightingale}},
  \bibnamefont{and} \bibinfo{author}{\bibfnamefont{M.}~\bibnamefont{den Nijs}},
  \bibinfo{journal}{Phys. Rev. Lett.} \textbf{\bibinfo{volume}{49}},
  \bibinfo{pages}{405} (\bibinfo{year}{1982}).

\bibitem[{\citenamefont{Niu et~al.}(1985)\citenamefont{Niu, Thouless, and
  Wu}}]{Thou2}
\bibinfo{author}{\bibfnamefont{Q.}~\bibnamefont{Niu}},
  \bibinfo{author}{\bibfnamefont{D.}~\bibnamefont{Thouless}}, \bibnamefont{and}
  \bibinfo{author}{\bibfnamefont{Y.-S.} \bibnamefont{Wu}},
  \bibinfo{journal}{Phys. Rev. B} \textbf{\bibinfo{volume}{31}},
  \bibinfo{pages}{3372} (\bibinfo{year}{1985}).

\bibitem[{\citenamefont{Kohmoto}(1985)}]{Khomo1}
\bibinfo{author}{\bibfnamefont{M.}~\bibnamefont{Kohmoto}},
  \bibinfo{journal}{Annals of Phys.} \textbf{\bibinfo{volume}{160}},
  \bibinfo{pages}{343} (\bibinfo{year}{1985}).

\bibitem[{\citenamefont{Avron and Seiler}(1985)}]{Avron1}
\bibinfo{author}{\bibfnamefont{J.~E.} \bibnamefont{Avron}} \bibnamefont{and}
  \bibinfo{author}{\bibfnamefont{R.}~\bibnamefont{Seiler}},
  \bibinfo{journal}{Phys. Rev. Lett.} \textbf{\bibinfo{volume}{54}},
  \bibinfo{pages}{259} (\bibinfo{year}{1985}).

\bibitem[{\citenamefont{Simon}(1983)}]{Simon1}
\bibinfo{author}{\bibfnamefont{D.}~\bibnamefont{Simon}},
  \bibinfo{journal}{Phys. Rev. Lett.} \textbf{\bibinfo{volume}{51}},
  \bibinfo{pages}{2167} (\bibinfo{year}{1983}).

\bibitem[{\citenamefont{Albrecht et~al.}(2001)\citenamefont{Albrecht, Smet, von
  Klitzing, Weiss, Umansky, and Schweizer}}]{Klit2}
\bibinfo{author}{\bibfnamefont{C.}~\bibnamefont{Albrecht}},
  \bibinfo{author}{\bibfnamefont{J.~H.} \bibnamefont{Smet}},
  \bibinfo{author}{\bibfnamefont{K.}~\bibnamefont{von Klitzing}},
  \bibinfo{author}{\bibfnamefont{D.}~\bibnamefont{Weiss}},
  \bibinfo{author}{\bibfnamefont{V.}~\bibnamefont{Umansky}}, \bibnamefont{and}
  \bibinfo{author}{\bibfnamefont{H.}~\bibnamefont{Schweizer}},
  \bibinfo{journal}{Phys. Rev. Lett.} \textbf{\bibinfo{volume}{86}},
  \bibinfo{pages}{147} (\bibinfo{year}{2001}).

\bibitem[{\citenamefont{Peierls}(1933)}]{Peierls1}
\bibinfo{author}{\bibfnamefont{R.}~\bibnamefont{Peierls}}, \bibinfo{journal}{Z.
  Phys.} \textbf{\bibinfo{volume}{80}}, \bibinfo{pages}{763}
  (\bibinfo{year}{1933}).

\bibitem[{\citenamefont{Harper}(1955)}]{Harper1}
\bibinfo{author}{\bibfnamefont{P.~G.} \bibnamefont{Harper}},
  \bibinfo{journal}{Proc. Phys. Soc.} \textbf{\bibinfo{volume}{A 68}},
  \bibinfo{pages}{874} (\bibinfo{year}{1955}).

\bibitem[{\citenamefont{Zak}(1964)}]{Zak1}
\bibinfo{author}{\bibfnamefont{J.}~\bibnamefont{Zak}}, \bibinfo{journal}{Phys.
  Rev.} \textbf{\bibinfo{volume}{134}}, \bibinfo{pages}{A1602}
  (\bibinfo{year}{1964}).

\bibitem[{\citenamefont{Azbel'}(1964)}]{Azbel1}
\bibinfo{author}{\bibfnamefont{M.~Y.} \bibnamefont{Azbel'}},
  \bibinfo{journal}{Sov. Phys. JETP} \textbf{\bibinfo{volume}{19}},
  \bibinfo{pages}{634} (\bibinfo{year}{1964}).

\bibitem[{\citenamefont{Rau}(1975)}]{Rauh1}
\bibinfo{author}{\bibfnamefont{A.}~\bibnamefont{Rau}}, \bibinfo{journal}{Phys.
  Status Solidi B} \textbf{\bibinfo{volume}{69}}, \bibinfo{pages}{K9}
  (\bibinfo{year}{1975}).

\bibitem[{\citenamefont{Dana and Zak}(1982)}]{Dana1}
\bibinfo{author}{\bibfnamefont{I.}~\bibnamefont{Dana}} \bibnamefont{and}
  \bibinfo{author}{\bibfnamefont{J.}~\bibnamefont{Zak}},
  \bibinfo{journal}{Phys. Rev. B} \textbf{\bibinfo{volume}{28}},
  \bibinfo{pages}{B811} (\bibinfo{year}{1982}).

\bibitem[{\citenamefont{Harper}(1991)}]{Harper3}
\bibinfo{author}{\bibfnamefont{P.~G.} \bibnamefont{Harper}},
  \bibinfo{journal}{J. Phys.: Condens. Matter} \textbf{\bibinfo{volume}{3}},
  \bibinfo{pages}{3047} (\bibinfo{year}{1991}).

\bibitem[{\citenamefont{Zak}(1997)}]{Zak2}
\bibinfo{author}{\bibfnamefont{J.}~\bibnamefont{Zak}}, \bibinfo{journal}{Phys.
  Rev. Lett.} \textbf{\bibinfo{volume}{79}}, \bibinfo{pages}{533}
  (\bibinfo{year}{1997}).

\bibitem[{\citenamefont{Hofstadter}(1976)}]{Hofta1}
\bibinfo{author}{\bibfnamefont{D.~R.} \bibnamefont{Hofstadter}},
  \bibinfo{journal}{Phys. Rev. B} \textbf{\bibinfo{volume}{14}},
  \bibinfo{pages}{2239} (\bibinfo{year}{1976}).

\bibitem[{\citenamefont{Kuhl and St{\"o}ckmann}(1998)}]{Kuhl1}
\bibinfo{author}{\bibfnamefont{U.}~\bibnamefont{Kuhl}} \bibnamefont{and}
  \bibinfo{author}{\bibfnamefont{H.~J.} \bibnamefont{St{\"o}ckmann}},
  \bibinfo{journal}{Phys. Rev. Lett.} \textbf{\bibinfo{volume}{80}},
  \bibinfo{pages}{3232} (\bibinfo{year}{1998}).

\bibitem[{\citenamefont{Richoux and Pagneux}(2002)}]{Richoux1}
\bibinfo{author}{\bibfnamefont{O.}~\bibnamefont{Richoux}} \bibnamefont{and}
  \bibinfo{author}{\bibfnamefont{V.}~\bibnamefont{Pagneux}},
  \bibinfo{journal}{Europhys. Lett.} \textbf{\bibinfo{volume}{59}},
  \bibinfo{pages}{34} (\bibinfo{year}{2002}).

\bibitem[{\citenamefont{Ashby and Miller}(1965)}]{Ashby1}
\bibinfo{author}{\bibfnamefont{N.}~\bibnamefont{Ashby}} \bibnamefont{and}
  \bibinfo{author}{\bibfnamefont{S.}~\bibnamefont{Miller}},
  \bibinfo{journal}{Phys. Rev. B} \textbf{\bibinfo{volume}{139}},
  \bibinfo{pages}{A428} (\bibinfo{year}{1965}).

\bibitem[{\citenamefont{Kunold and Torres}(2000)}]{Kunold1}
\bibinfo{author}{\bibfnamefont{A.}~\bibnamefont{Kunold}} \bibnamefont{and}
  \bibinfo{author}{\bibfnamefont{M.}~\bibnamefont{Torres}},
  \bibinfo{journal}{Phys. Rev. B} \textbf{\bibinfo{volume}{61}},
  \bibinfo{pages}{9879} (\bibinfo{year}{2000}).

\bibitem[{\citenamefont{Hadjioannou and Sarlis}(1996)}]{Hadji1}
\bibinfo{author}{\bibfnamefont{F.~T.} \bibnamefont{Hadjioannou}}
  \bibnamefont{and} \bibinfo{author}{\bibfnamefont{N.~V.}
  \bibnamefont{Sarlis}}, \bibinfo{journal}{Phys. Rev. B}
  \textbf{\bibinfo{volume}{54}}, \bibinfo{pages}{5334} (\bibinfo{year}{1996}).

\bibitem[{\citenamefont{Hadjioannou and Sarlis}(1997)}]{Hadji2}
\bibinfo{author}{\bibfnamefont{F.~T.} \bibnamefont{Hadjioannou}}
  \bibnamefont{and} \bibinfo{author}{\bibfnamefont{N.~V.}
  \bibnamefont{Sarlis}}, \bibinfo{journal}{Phys. Rev. B}
  \textbf{\bibinfo{volume}{56}}, \bibinfo{pages}{9406} (\bibinfo{year}{1997}).

\bibitem[{\citenamefont{Zak}(1993)}]{Zak3}
\bibinfo{author}{\bibfnamefont{J.}~\bibnamefont{Zak}}, \bibinfo{journal}{Phys.
  Rev. Lett.} \textbf{\bibinfo{volume}{71}}, \bibinfo{pages}{2623}
  (\bibinfo{year}{1993}).

\bibitem[{\citenamefont{Zak}(1967)}]{Zak4}
\bibinfo{author}{\bibfnamefont{J.}~\bibnamefont{Zak}}, \bibinfo{journal}{Phys.
  Rev. Lett.} \textbf{\bibinfo{volume}{19}}, \bibinfo{pages}{1385}
  (\bibinfo{year}{1967}).

\bibitem[{\citenamefont{Moshinsky and Quesne}(1971)}]{Moch1}
\bibinfo{author}{\bibfnamefont{M.}~\bibnamefont{Moshinsky}} \bibnamefont{and}
  \bibinfo{author}{\bibfnamefont{C.}~\bibnamefont{Quesne}},
  \bibinfo{journal}{Journal of Mathematical Physics}
  \textbf{\bibinfo{volume}{12}}, \bibinfo{pages}{1772} (\bibinfo{year}{1971}).

\bibitem[{\citenamefont{Risken}(1984)}]{Risken1}
\bibinfo{author}{\bibfnamefont{H.}~\bibnamefont{Risken}},
  \emph{\bibinfo{title}{The Fokker-Planck Equation-Methods of Solution and
  Applications}} (\bibinfo{publisher}{Springer-Verlag},
  \bibinfo{address}{Berlin}, \bibinfo{year}{1984}), vol.~\bibinfo{volume}{18}
  of \emph{\bibinfo{series}{Springer Series in Synergetics}},
  chap.~\bibinfo{chapter}{9}.

\bibitem[{\citenamefont{Petschel and Geisel}(1993)}]{Geisel1}
\bibinfo{author}{\bibfnamefont{G.}~\bibnamefont{Petschel}} \bibnamefont{and}
  \bibinfo{author}{\bibfnamefont{T.}~\bibnamefont{Geisel}},
  \bibinfo{journal}{Phys. Rev. Lett.} \textbf{\bibinfo{volume}{71}},
  \bibinfo{pages}{239} (\bibinfo{year}{1993}).

\bibitem[{\citenamefont{Haken}(1976)}]{Haken1}
\bibinfo{author}{\bibfnamefont{H.}~\bibnamefont{Haken}},
  \emph{\bibinfo{title}{Quantum Field Theory of Solids}}
  (\bibinfo{publisher}{North-Holland}, \bibinfo{address}{London},
  \bibinfo{year}{1976}), chap.~\bibinfo{chapter}{14}.

\bibitem[{\citenamefont{Callaway}(1974)}]{Callaway}
\bibinfo{author}{\bibfnamefont{J.}~\bibnamefont{Callaway}},
  \emph{\bibinfo{title}{Quantum Theory of the Solid State}}
  (\bibinfo{publisher}{Academic Press, Inc.}, \bibinfo{year}{1974}),
  chap.~\bibinfo{chapter}{6}.

\bibitem[{\citenamefont{Chang and Niu}(1996)}]{MingChe1}
\bibinfo{author}{\bibfnamefont{M.-C.} \bibnamefont{Chang}} \bibnamefont{and}
  \bibinfo{author}{\bibfnamefont{Q.}~\bibnamefont{Niu}},
  \bibinfo{journal}{Phys. Rev. B} \textbf{\bibinfo{volume}{53}},
  \bibinfo{pages}{7010} (\bibinfo{year}{1996}).

\bibitem[{\citenamefont{Kunold}(2003)}]{Kunold2}
\bibinfo{author}{\bibfnamefont{A.}~\bibnamefont{Kunold}}, Ph.D. thesis,
  \bibinfo{school}{Universidad Nacional Aut\'onoma de M\'exico, Instituto de
  F\'{\i}sica}, \bibinfo{address}{Apartado Postal 20-364, M\'exico D.F. 01000,
  M\'exico} (\bibinfo{year}{2003}).

\end{thebibliography}

\end{document}